\documentclass[11pt]{article}
\usepackage{latexsym}
\usepackage{amsmath}
\usepackage{verbatim}
\usepackage{mathrsfs}

\usepackage{amssymb,amsmath}
\usepackage{amsthm}
\usepackage{amscd}
\usepackage{amssymb}
\usepackage{amsfonts}
\usepackage{url}
\usepackage{slashed}
\usepackage[latin1]{inputenc}
\usepackage[english]{babel}

\usepackage{epsfig}
\usepackage{subfigure}

 \csname
@addtoreset\endcsname{equation}{section}

\def\gz0{\gamma^{0}}

\def\scs#1{\section{\sc #1}}

\def\g{\gamma}

\def\vf{\varphi}

\def\be{\begin{equation}}
\def\ee{\end{equation}}
\def\bs{\begin{split}}
\def\es{\end{split}}
\def\bea{\begin{eqnarray}}
\def\eea{\end{eqnarray}}
\def\ba{\begin{array}}
\def\ea{\end{array}}
\def\bec{\begin{center}}
\def\ec{\end{center}}
\def\ba{\begin{align}}
\def\ena{\end{align}}

\def\12{\frac{1}{2}}

\def\dag{\dagger}

\newcommand{\equ}[2]{\begin{equation}\label{#1}{#2}\end{equation}}

\begin{document}

\begin{flushright}
{\today}\\
CPHT-RR 054.0711\\
LPTENS-11-26\\
CERN-PH-TH/2011-324
\end{flushright}

\vspace{15pt}

\begin{center}
{\Large\sc CMB Imprints of a Pre-Inflationary Climbing Phase}\\
\vspace{25pt}
{\sc E.~Dudas${}^{\; a,\;b,\;c}$, N.~Kitazawa${}^{\; d}$, S.~P.~Patil${}^{\; b,\; e}$ and A.~Sagnotti$^{\; f}$}\\[15pt]

{${}^a$\sl\small Theory Group, Physics Department,
CERN CH--1211\\
Geneva 23 \ SWITZERLAND\\}\vspace{10pt}

{${}^b$\sl\small Centre de Physique Th\'eorique,
\'Ecole Polyt\'echnique,
F-91128 Palaiseau \ FRANCE\\}e-mail: {\small \it

dudas@cpht.polytechnique.fr;
patil@cpht.polytechnique.fr}\vspace{10pt}

{${}^c$\sl\small LPT (UMR CNRS 8627), Bat 210, Universit\'e de Paris-Sud, F-91405 Orsay Cedex \ FRANCE}\vspace{10pt}

{${}^d$\sl\small Department of Physics, Tokyo Metropolitan University\\
Hachioji, Tokyo, 192-0397 \ JAPAN
\\ }e-mail: {\small \it
kitazawa@phys.se.tmu.ac.jp}\vspace{10pt}

{${}^e$\sl\small Laboratoire de Physique Th\'eorique, \'Ecole Normale Sup\'erieure\\
24 Rue Lhomond, F-75005 Paris \ FRANCE
\\ }\vspace{10pt}

{${}^f$\sl\small
Scuola Normale Superiore and INFN\\
Piazza dei Cavalieri, 7, I-56126 Pisa \ ITALY \\}
e-mail: {\small \it sagnotti@sns.it}\vspace{10pt}
\vspace{12pt}

{\sc\large Abstract}
\end{center}
{We discuss the implications for cosmic microwave background (CMB) observables, of a class of pre--inflationary dynamics suggested by string models where SUSY is broken due to the presence of D--branes and orientifolds preserving incompatible portions of it. In these models the would--be inflaton is forced to emerge from the initial singularity climbing up a mild exponential potential, until it bounces against a steep exponential potential of ``brane SUSY breaking'' scenarios, and as a result the ensuing descent gives rise to an inflationary epoch that begins when the system is still well off its eventual attractor. If a pre-inflationary climbing phase of this type had occurred within 6-7 $e$--folds of the horizon exit for the largest observable wavelengths, displacement off the attractor and initial--state effects would conspire to suppress power in the primordial scalar spectrum, enhancing it in the tensor spectrum and typically superposing oscillations on both. We investigate these imprints on CMB observables over a range of parameters, examine their statistical significance, and provide a semi--analytic rationale for our results. It is tempting to ascribe at least part of the large-angle anomalies in the CMB to pre--inflationary dynamics of this type.}

\setcounter{page}{1}
\pagebreak
{\linespread{0.75}\tableofcontents}
\newpage


\scs{Introduction}\label{sec:intro}

An epoch of primordial inflation \cite{inflation} is widely accepted to have premised the Big Bang, since it readily accounts for the flatness, homogeneity, isotropy and causal structure of the initially thermalized plasma, as well as for the apparent absence of dangerous topological relics in our Universe. In its simplest realizations, inflation is also remarkably successful in accounting for the observed CMB anisotropies, and provides a quantum mechanical origin for the primordial inhomogeneities around which we believe that all visible structure formed.

In order to solve the flatness problem, inflation must have lasted for a certain number of $e$--folds, with a lower bound that depends on the putative reheating temperature of the Universe (at least 58 $e$--folds for GUT--scale reheating). In toy models where the inflaton is treated as a classical scalar field minimally coupled to gravity, substantially more than this minimal $e$--folding can readily be attained with fine--tuned potentials or large field excursions \cite{ll}. Matters are markedly different, however, if one demands that inflation result from an effective field theory (EFT) where the inflaton is an effectively light degree of freedom, since loop corrections tend to make long periods of inflation unnatural. If the effective potential, rather than the classical potential, is to be responsible for the phenomenon \cite{eft1}, various tunings are in fact to be made or various symmetries are to be invoked, which are furthermore required to survive the UV completion of the theory. One of the most challenging issues for model builders is perhaps the ``eta problem", wherein corrections to the inflaton mass of order $H^2$, where $H$ is the Hubble parameter, are readily generated by operators of dimension six in the EFT, while the inflaton mass scale ought to be far lower for successful inflation (see \cite{bg} for a discussion of the problem and a proposal to evade it). In spite of many suggestions to bypass it in specific contexts (see also \cite{eta}), the eta problem seems to persist in generality \cite{etaprob}\footnote{See also \cite{palma} for a discussion of the eta problem as it relates to uplifting in supergravity theories.}.

Little is known about the actual EFT in which inflation is to be embedded, and as we have stressed, sustaining it for the needed duration appears to be a delicate proposition in realistic constructions. In low--energy supergravity models inspired by String Theory, the challenge is to arrange for the inflaton to undergo slow roll in a region of the potential that can sustain it without overshooting \cite{overshoot}, while in general EFT treatments \cite{eft1,infl_EFT} the challenge is to arrange for effective potentials leading to inflaton masses that are much smaller than the Hubble scale. All these requirements are ever more challenging if one requires direct couplings to a matter sector that contains the Standard Model, which is needed if reheating is to occur efficiently\footnote{See \cite{cicoli} for interesting observational consequences that could result in the context of large volume compactifications.}. In light of these considerations, it behooves us (as it has behooved other authors \cite{just}) to entertain the possibility that inflation did not last that much longer than is needed to solve the flatness problem.

Long--wavelength distortions of CMB observables are obvious corollaries of short inflation. Although we are limited by cosmic variance at the largest angular scales, some statistically significant imprints of a possible pre--inflationary phase could in fact persist in the sky if inflation did not last more than 6--7 $e$--folds longer than it had to. For instance, a single pre--inflationary particle could have sourced ring--like structures in the CMB \cite{pip, pip_sig}\footnote{The many--particle superposition of this effect would bring about an excess of power (\emph{i.e.} a feature) at long wavelengths in the power spectrum.}. Alternatively, a non-vanishing CMB dipole sourced by super--horizon inhomogeneities that were not completely washed away, although insignificant for the CMB power spectrum, could have had sizable consequences for the large--scale structure of the Universe. To wit, there may be evidence for a large--scale bulk flow \cite{enhanced} that is in significant discrepancy with the predictions of $\Lambda$CDM models, which could readily be explained if the CMB possessed an intrinsic dipole moment \cite{dipole}. Taken together with statistically significant observations \cite{low_power} of the lack of power at large angles in the angular correlation function (as opposed to the angular power spectrum)\footnote{Taking into account the observed alignments at low multipoles, the chances that such large--angle anomalies would be produced by a concordance model has been estimated by the authors of \cite{low_power} as $1$ in $10^6$.}, this suggests that we may be seeing something other than the vanilla initial conditions generated by a phase of long inflation.

With this in mind we explore here the consequences for CMB observables, of a genre of pre-inflationary dynamics \cite{dks} that presents itself in String Theory \cite{stringtheory}, in orientifold vacua \cite{orientifolds} where supersymmetry (SUSY) is broken by \emph{classically stable} combinations of (anti) D-branes and orientifolds \cite{bsb}.
The peculiarity of these models is that a scalar field is typically forced to emerge from the initial singularity climbing up an exponential potential, so that it can subsequently inject a slow--roll phase that begins off the eventual attractor \cite{dks}~\footnote{We thank E.~Kiritsis for calling to our attention some previous work on exponential potentials when the results of \cite{dks} were presented at ``String Phenomenology 2010''. A vast previous literature actually includes \cite{attractor,halliwell,pli,dm2,exponential,townsend} and is mostly devoted to the asymptotic behavior of similar classes of models even with more scalar fields, but Russo discussed in \cite{townsend} the classical solutions that we about to review in Section~\ref{sec:climbing}.}. As we shall see, aside from providing a natural mechanism to initially displace the inflaton up along its potential (so that the origin of its initial potential energy is accounted for), this climbing phase and the transients associated with it conspire to modify the power spectrum at long wavelengths. We find suppression of power at the largest scales for the scalar spectrum, (mild) enhancement for the tensor spectrum, with superposed oscillations at long wavelengths in both cases. We anticipate that a better fit to CMB observations may result and account at least for part of the large--angle anomaly previously discussed.

The relevant mechanism, originally termed ``brane SUSY breaking'' \cite{bsb}, results from the presence of (anti)branes and orientifolds preserving incompatible fractions of the original SUSY, which is dictated by charge constraints in some orientifold vacua of String Theory. This gives rise to tree--level exponential potentials, whose lack of critical points would bring about in a flat background, tadpoles signalling the need for complicated resummations \cite{fs,wv}. In a cosmological setting, however, the same potentials have the intriguing consequence that a scalar field can be \emph{forced} to emerge from the initial singularity while climbing up. As discussed in \cite{dks}, this is already the case in the simplest instance of ``brane SUSY breaking'', the ten--dimensional Sugimoto model \cite{sugimoto}, where the initial climbing can even inject an eventual slow--roll phase driven by a milder exponential \cite{dms} of non--perturbative origin. Similar phenomena take place in the richer four--dimensional KKLT scenario \cite{kklt}, whose ``uplift'' may also be ascribed to the phenomenon of ``brane SUSY breaking'' and gives rise again to the climbing behavior. From this viewpoint, KKLT--like scenarios are also interesting since, as shown in \cite{dks}, the combined effects of a climbing phase and cosmological friction could have trapped some moduli in metastable vacua right after the initial singularity. A natural objection to this type of analysis is that even restricting the attention to tree--level potentials, which are expected to dominate at weak coupling in String Theory, curvature corrections near the initial singularity could invalidate the climbing mechanism. We are well aware of the problem, and we do not have as yet a truly satisfactory answer to this important question. Still, we are persuaded that a mechanism that is capable of naturally injecting inflation deserves to be investigated further, and it is conceivable that our setup will survive the relevant corrections at least in special classes of models. Moreover, the detailed consequences for CMB observables that we are about to describe depend largely upon the dynamics of the climbing inflaton at later times, when it idles about its eventual attractor. Although clearly embodied by specific string--inspired constructions where a single minimally--coupled scalar field is subject to the relatively simple class of potentials
\be
V(\phi) \, = \, \frac{M^2}{2\kappa^2}\  \left( e^{\, \, \sqrt{6} \, \phi} \ + \ e^{\, \, \sqrt{6} \, \gamma\, \phi} \right)  \ , \label{exppotintro}
\ee
where a ``hard'' exponential brought about by brane SUSY breaking accompanies a slow--roll term (with $0<\gamma<1/\sqrt{3}$) that dominates the eventual inflationary dynamics, this type of phenomenon ought to be relevant in a wider context. As we shall see, the scalar is forced to emerge from the initial singularity climbing up the mild portion of the potential, so that at some point it experiences the hard exponential of String Theory, essentially bouncing against it. The ensuing dynamics equilibrates rather slowly to the eventual attractor, and this brings about a sizable infrared suppression of the CMB power spectrum. Our result is not completely new, since suppression of low CMB multipoles was also related in \cite{destri} to a pre--inflationary fast roll in double--well potentials, and more recently similar effects were also considered in \cite{just}\footnote{Recently we also became aware of \cite{piao}, which refers however to a different context, a pre--inflationary bounce motivated by the pre--Big Bang scenarios of \cite{pbb}.}. As we shall see, however, while a suppression is generic in models with an initial singularity, it is particularly sizable for those scalars {\it that are initially forced to climb until they bounce off a hard exponential ``wall''}, with statistically significant consequences over a larger range of lower multipole moments. Furthermore, we note that the potential (\ref{exppotintro}) represents a generic parametrization of brane SUSY breaking scenarios that feed and sustain inflation initially. Since our study is only concerned with the effects on long--wavelength CMB observables, we can safely extricate our analysis from the effects of terms in the effective potential that become relevant as smaller scales exit the horizon. Hence the question of how inflation ends, while important for ensuring a viable phenomenology, represents a separate issue that does not affect our conclusions.

We can summarize our findings as having uncovered:
\begin{itemize}
\item
a statistically significant suppression of angular power over a range of lower multipoles,
\item
which reflects pre-inflationary dynamics that settled onto slow--roll up to several $e$--folds before the present Hubble scale exited the horizon.
\end{itemize}

Our construction was motivated by a class of string constructions:
\begin{itemize}
\item
where the breaking of SUSY is closely tied to the string scale;
\item
which provide naturally the very impetus for inflation to begin;
\item
which motivate a new class of phenomena -- starting inflation off the attractor.
\end{itemize}

The paper opens in Section~\ref{sec:climbing} with a brief review of how a single minimally--coupled scalar field with an exponential potential of large enough slope is forced to emerge from the initial singularity while climbing up. We leave out a discussion of the more complicated KKLT--type constructions (for which the reader is referred to \cite{dks}), since the ensuing analysis of CMB perturbations is restricted, for the sake of simplicity and predictivity, to models whose cosmological evolution is dominated by a single scalar field. A closer look at the dynamics of this scalar field, which is presented in the following sections paying attention to both pre--inflationary and slow--roll regimes, shows that \emph{inflation can effectively begin long before the solution comes close to the eventual attractor}. In Section~\ref{sec:perturbations} we begin to analyze the behavior of correlation functions in this class of backgrounds, starting from a perturbative treatment of the corrections to CMB observables induced by inflating off the attractor, and in Section~\ref{sec:powerspectrum} we review some key properties of scalar and tensor power spectra. In Section~\ref{sec:numerics} we numerically evaluate these power spectra for various choices of parameters of the climbing/inflationary potential, present an understanding of our results in terms of spectral running, and discuss the prospects for offering a better fit to observations. In Section~\ref{sec:qma} we elaborate upon a general, albeit semi--quantitative, WKB picture of the observable consequences of a ``hard'' exponential and the associated climbing phase,
relying on some similarities between the Mukhanov-Sasaki and Schr\"odinger problems.  Section~\ref{sec:conclusions} contains our conclusions. Finally, Appendix~\ref{a1} summarizes general features of cosmological perturbations, Appendix~\ref{a3} describes an analytic model for scalar and tensor power spectra and Appendix~\ref{a2} addresses the issue of initial--state effects on their long--wavelength tails. In the paper we shall refer to three types of time derivatives, with respect to a parametric time $\tau$ that makes it possible to obtain some useful analytic information on the system, with respect to the cosmological time $t$ and with respect to the conformal time $\eta$. Following a standard practice in Cosmology, the latter two will denoted by ``dots'' and ``primes'' respectively, while the first type of derivatives will be indicated explicitly. These conventions differ from those of \cite{dks}, where ``dots'' referred to $\tau$--derivatives.

\vskip 24pt


\scs{Climbing scalars and String Theory}\label{sec:climbing}

Let us begin with a brief review of some results of \cite{dks}. Our analysis rests on a class of low--energy four--dimensional actions of the type\footnote{In \cite{dks} we were abiding to standard conventions in String Theory. Here we change slightly the Lagrangian, in order to work with a canonically normalized scalar field.}
\be
S \ = \ \int d^{4} x \, \sqrt{-g}\,
    \left[ \frac{1}{{2\kappa^2}} \, R \, - \, \frac{1}{2}\ (\partial \phi)^2
            \, - \, V(\phi) \, + \, \ldots \right]\, , \label{action_phi}
\ee
for a single scalar field coupled to gravity, where for notational simplicity we readily turn to Planck units, setting from now on $\kappa^{\,2}=8 \pi G_N=1$. One can study an interesting class of cosmological solutions letting
\be
ds^2 \, =\, - \, e^{\,
2B(\xi)}\, d\xi^2 \, + \, e^{\, 2A(\xi)} \, d{\bf x} \cdot d{\bf x} \ ,
\qquad
\phi = \phi(\xi) \ , \label{metric}
\ee
and making the convenient gauge choice \cite{halliwell,dm2,townsend}
\be
V(\phi) \ e^{2B} \, = \, \frac{M^2}{2} \ , \label{gauge}
\ee
where $M$ is a mass scale related to the potential $V(\phi)$ \footnote{In general units, this gauge choice would involve the dimension--four ratio $\frac{M^2}{2\kappa^2}$.}. With the redefinitions
\be
\tau \,=\, M\, \sqrt{\frac{3}{2}} \ \xi \ , \qquad \vf \, = \, \sqrt{\frac{3}{2}} \ \phi \ , \qquad {\cal A} \,=\, 3 \, A\
, \label{defs} \ee
and confining our attention to an expanding Universe, so that
\be \frac{d{\cal A}}{d\tau} = \sqrt{1 \, +\, \left(\frac{d \vf}{d\tau}\right)^2}  \ , \ee
one is thus led to recast the usual dynamical equation
\be
{\ddot \phi}\ + \ 3\, H \, {\dot \phi} \ + \ V_\phi \ = \ 0 \ , \label{eqphit}
\ee
where ``dots'' denote derivatives with respect to the cosmological time and
\be
H \ = \ \frac{1}{\sqrt{6}} \, \sqrt{2V \, +\, {\dot \phi}^{\,2}} \ , \label{Hphit}
\ee
into the convenient form
\be\label{eqphi} \frac{d^{\,2} \vf}{d\tau^{\,2}} \, + \, \frac{d \vf}{d\tau} \, \sqrt{1\,+\, \left(\frac{d \vf}{d\tau}\right)^2}
\, +\, \left[ 1\,+\, \left(\frac{d \vf}{d\tau}\right)^2 \,\right]\ \frac{V_\vf}{2V} \, =\, 0\,  \ee
that only involves the logarithmic derivative of the potential $V$, and thus anticipates the key role of exponential potentials for these systems.

Potentials of the type
\be
V \, = \, \frac{M^2}{2}\  \left( e^{\, 2 \, \vf} \ + \ e^{\, 2 \, \gamma\, \vf} \right) \, = \, \frac{M^2}{2}\  \left( e^{\, \, \sqrt{6} \, \phi} \ + \ e^{\, \, \sqrt{6} \, \gamma\, \phi} \right) \ , \label{exppot2}
\ee
depending on a pair of exponential functions, with $0<\gamma<1$ and where a positive relative coefficient could be absorbed by a shift of $\vf$, are naturally present in String Theory \cite{stringtheory} and particularly so in orientifold models \cite{orientifolds} with ``brane SUSY breaking'' \cite{bsb}. In these models SUSY, broken \emph{at the string scale} because charge conservation requires that (anti) D--branes and orientifold planes respecting incompatible portions of it be present, is non--linearly realized \emph{\`a la} Volkov--Akulov \cite{nlsusy} in the low--energy supergravity \cite{dm1}. In flat space exponential potentials of this type are somehow a nuisance, since their lack of critical points brings about tadpoles and thus the need for complicated vacuum redefinitions \cite{fs,wv}, but in cosmological settings they give rise to some interesting effects that we are about to recall.

This type of scenario finds its simplest realization in the ten--dimensional Sugimoto model \cite{sugimoto}, where $\vf$ can be related to the dilaton. In this case the first term in \eqref{exppot2} reflects (projective) disk contributions accounting for the D9 anti-branes and the O9 planes present in the perturbative vacuum while the second accounts for the non-BPS D3 brane that was identified in \cite{dms} following \cite{sen}. The setting is also relevant, however, for the four--dimensional KKLT scenario \cite{kklt}, where as stressed in \cite{dks} F--term ``uplift'' potentials draw their origin from the first term of eq.~\eqref{exppot2} while $\vf$ can be related, along the lines of \cite{witten85}, to an internal breathing mode. We shall keep this richer scenario in mind in the ensuing analysis, but in this paper we restrict our attention to the single--field setting. Taking into account the KKLT axion would complicate the analysis of cosmological perturbations, while as shown in \cite{dks} the early climbing behavior of the KKLT system is well captured by the scalar field alone.

The single--scalar model affords interesting exact solutions for the simpler class of potentials
\be
V \, = \, \frac{M^2}{2} \ e^{\, 2 \, \gamma\, \vf} \ , \label{exppot1}
\ee
whose analysis has a long history \cite{attractor,halliwell,pli,dm2,exponential,townsend}. To earlier results on general attractor trajectories and to the more complete analysis in \cite{townsend}, in particular in Russo's paper, \cite{dks} added more recently the recognition of some general implications of the climbing phenomenon, to which we now turn our attention.

For the potential \eqref{exppot1}, eq.~\eqref{eqphi} reduces to
\be\label{eqphigamma} \frac{d^{\,2} \vf}{d\tau^{\,2}} \, + \, \frac{d \vf}{d\tau} \, \sqrt{1\,+\, \left(\frac{d \vf}{d\tau}\right)^2}
\, +\, \gamma\, \left[ 1\,+\, \left(\frac{d \vf}{d\tau}\right)^2 \,\right]\, \, =\, 0\,  \ee
that for $0< \g < 1$ admits the two distinct exact solutions $(0 < \tau < \infty)$
\be
\frac{d \vf}{d\tau} \, = \, \frac{1}{2} \left[ \sqrt{\frac{1\,-\, \g}{1\,+\, \g}}\, \tanh \left( \frac{\tau}{2}\ \sqrt{1\,-\, \g^{\,2}} \,\right) \ - \ \sqrt{\frac{1\,+\, \g}{1\,-\, \g}}\, \coth \left( \frac{\tau}{2}\ \sqrt{1\,-\, \g^{\,2}}\, \right)\right] \label{descending_phi}
\ee
and
\be
\frac{d \vf}{d\tau} \, = \, \frac{1}{2} \left[ \sqrt{\frac{1\,-\, \g}{1\,+\, \g}}\, \coth \left( \frac{\tau}{2}\ \sqrt{1\,-\, \g^{\,2}}\, \right) \ - \ \sqrt{\frac{1\,+\, \g}{1\,-\, \g}}\, \tanh \left( \frac{\tau}{2}\ \sqrt{1\,-\, \g^{\,2}}\, \right)\right]\ . \label{climbing_phi}
\ee
Eq.~(\ref{descending_phi}) describes a scalar field that emerges from the Big Bang \emph{climbing down} the exponential potential, whereas the field $\vf$ in (\ref{climbing_phi}) initially \emph{climbs up}, then reverts its motion and eventually climbs down. In both cases the scalar field eventually approaches in terms of the ``parametric time'' $\tau$ the ``limiting speed''
\be
\frac{d \vf}{d\tau} \, = \, - \ \frac{\g}{\sqrt{1\, - \, \g^{\, 2}}} \ . \label{limspeed}
\ee
Notice that this limiting speed can be derived directly from eq.~\eqref{eqphigamma}, as for a simpler analogue of this system -- a Newtonian particle in a viscous medium. It ceases to exist when $\g$ approaches one from below, and when this critical value is overcome the two classes of solutions in eqs.~\eqref{descending_phi} and \eqref{climbing_phi} leave way to a single class that describes a climbing scalar, with
\be \frac{d \vf}{d\tau} \, = \, \frac{1}{2} \left[ \sqrt{\frac{\g \,-\, 1}{\g \,+\, 1}}\, \cot \left( \frac{\tau}{2}\ \sqrt{\g^{\,2} \,-\, 1}\, \right) \ - \ \sqrt{\frac{\g\,+\, 1}{\g \,-\, 1}}\, \tan \left( \frac{\tau}{2}\ \sqrt{\g^{\,2} \,-\, 1}\, \right)\right]\ , \label{superclimbing_phi}
\ee
where now $0< \tau < \frac{\pi}{\sqrt{\g^{\,2} \,-\, 1}}$  \cite{dks}.

Considerations of this type capture important features of the two--term potential \eqref{exppot2}, whose dynamics is interestingly richer. If $0<\gamma<1$, the asymptotic behavior for large positive $\vf$ is clearly dominated by the first term, which makes an initial climbing phase unavoidable, while the descent toward large negative values of $\vf$ is dominated by the second term, so that $\vf$ is bound to approach asymptotically the limiting speed \eqref{limspeed}. Actually, when $\vf$ comes close to terminating its ascent, the steep first term in eq.~\eqref{exppot2} drives it to invert its motion somewhat abruptly, thus paying off less energy to cosmological friction than it would in the milder single--term potential \eqref{exppot1}. As a result, its descent proceeds readily at $\tau$--speeds larger than \eqref{limspeed}, which is thus approached from above. This qualitative feature will play an important role in Sections~\ref{sec:perturbations} and \ref{sec:qma}. Finally, as discussed in \cite{dks}, if $0<\g <\frac{1}{\sqrt{3}}$ the limiting speed \eqref{limspeed} translates into slow roll, so that even in the simplest manifestation of brane SUSY breaking, a one--field system with potential \eqref{exppot2} and $\g= \frac{1}{2}$, the initial climbing is bound to inject an inflationary phase.

An important exact solution of the single--exponential model, the Lucchin--Matarrese (LM) attractor \cite{attractor}, exists for $0<\g<1$ and describes a scalar field that proceeds at the limiting speed \eqref{limspeed}, which is also the large--$\tau$ behavior of generic solutions. In this case
\be
\vf \, = \, \vf_0\ -\ \frac{\g \, \tau}{\sqrt{1\,-\, \g^{\,2}}} \ , \qquad {\cal A}  \, = \, \frac{\tau}{\sqrt{1\,-\, \g^{\,2}}} \ , \label{LMtau}
\ee
where in the second expression we are omitting an additive constant that could be absorbed rescaling the spatial coordinates. In terms of the \emph{cosmological time} $t$, defined by
\be
dt \, = \, e^B d \xi \,=\, \frac{1}{\sqrt{3\, V(\phi)}} \ d\tau \label{cosmict}
\ee
so that the metric \eqref{metric} takes the form
\be
ds^2 \, =\, - \, dt^2 \ +\ e^{\, \frac{2}{3}\, {\cal A}(\xi)} \, d{\bf x} \cdot d{\bf x}\ ,
\ee
eqs.~\eqref{LMtau} become
\be
\vf \, = \, - \, \frac{1}{\g} \ \ln \left[ \sqrt{\frac{3}{2}} \ \frac{\g^{\,2}\, M\, t}{\sqrt{1\,-\, \g^{\,2}}}\right] \ , \qquad {\cal A}  \, = \, \frac{1}{\gamma}\ \vf_0 \ + \ \frac{1}{\g^{\,2}} \ \ln \left[ \sqrt{\frac{3}{2}} \ \frac{\g^{\,2}\, M\, t}{\sqrt{1\,-\, \g^{\,2}}}\right] \ ,
\ee
and therefore the LM metric exhibits the power--like behavior
\be
ds^2 \, = \, -\, d t^2 \, +\, \left(\sqrt{\frac{3}{2}}\ \frac{M \, \gamma^{\,2}\, t}{\sqrt{1-\gamma^{\,2}}} \ e^{\,\gamma\,\vf_0\,} \right)^\frac{2}{3\,\gamma^{\,2}} \, d{\bf x} \cdot d{\bf x}\ . \label{metric_cosmfLM}
\ee
The LM attractor \cite{attractor,pli,exponential} thus describes a scalar field that actually slows down indefinitely in terms of the cosmological time $t$. This type of attractor solution plays a key role in the analysis of the CMB, and our task in the next sections will be to investigate how departures from it implied by the string--motivated picture of a scalar that climbs up the potential at early times can affect CMB observables. In a similar fashion, near the initial singularity, as the scalar climbs up the mild portion of the potential \eqref{exppot2} depending on $\gamma$,
\be
\vf \, \sim \, {\cal A} \, \sim \, \frac{1}{1\,+\, \gamma} \ \log \tau \, +\, \vf_0 \ ,
\ee
so that eq.~\eqref{cosmict} implies that $t \sim \tau^\frac{1}{1+\gamma}$ and
\be
ds^2 \, \sim \, -\, d t^2 \, +\, \left(\sqrt{\frac{2}{3}}\ \frac{M t}{{1+\gamma}} \ e^{\,(1\,+\,\gamma)\vf_0\,} \right)^\frac{2}{3} \, d{\bf x} \cdot d{\bf x}\ . \label{cosmic_climbing}
\ee

It is also useful to introduce the \emph{conformal time} $\eta$, such that
\be
d t \, = \, e^{\, \frac{1}{3}\,{\cal A(\xi)}} \ d \eta \ , \label{conf_time}
\ee
which is conventionally defined so that it approaches zero from negative values at late epochs, in terms of which the metric takes the form
\be
ds^2 \, =\, a^2(\eta) \, \left( \, -\, d \eta^2 \, +\, d{\bf x} \cdot d{\bf x} \right) \, , \label{metric_conf}
\ee
where the conformal scale factor is
\be
a(\eta) \, = \, e^{\, \frac{1}{3}\, {\cal A}(\eta)} \ . \label{conf_factor}
\ee
In particular, for the LM attractor
\be
\eta \, = \, -\, \frac{\sqrt{6(1\,-\, \gamma^{\,2})}}{M(1\,-\,3\, \gamma^{\,2})} \ \exp\left( \,-\, \tau\, \frac{1\,-\, 3\, \gamma^{\,2}}{3\, \sqrt{1\,-\, \gamma^{\,2}}}\ - \ \gamma \ \vf_0  \right) \ ,
\ee
so that
\be
\begin{split}
ds^{\,2} \, &=\, \left( \frac{\sqrt{6(1\,-\, \gamma^{\,2})}}{M(1\,-\,3\, \gamma^{\,2})} \ e^{\,-\, \gamma\, \vf_0} \right)^\frac{2}{1\,-\,3\,\gamma^{\,2}} \left( \,-\, \eta \right)^{\,-\, \frac{2}{1\,-\,3\,\gamma^{\,2}}} \ \bigg( \, -\, d \eta^2 \, +\, d{\bf x} \cdot d{\bf x} \bigg) \ , \\ \label{metric_confLM}
\vf \, &= \, \frac{\vf_{\,0}}{1\ - \ 3\, \g^{\,2}}\ + \ \frac{3\, \g}{1\ - \ 3\, \g^{\,2}} \ \log\left[ \frac{M(1\,-\,3\, \gamma^{\,2})}{\sqrt{6(1\,-\, \gamma^{\,2})}} \ \big(-\, \eta \big) \right] \ .
\end{split}
\ee

We can now move on to correlation functions of gauge invariant perturbations around the background climbing solution of the potential \eqref{exppot2}. We begin with some basic aspects of cosmological perturbation theory that are needed for our analysis, which are further elaborated upon in Appendix~\ref{a1}, and then turn our attention to the so--called scalar and tensor spectra for the CMB anisotropies. These primordial spectra will form the basis of our comparison with observations in Section~\ref{sec:numerics}, which we shall also reconsider in a WKB setting in Section~\ref{sec:qma}.

\vskip 24pt


\scs{Correlation functions off the attractor}\label{sec:perturbations}


The models of the preceding section provide a natural mechanism to give inflation its initial impetus. A phase of pre--inflationary climbing (beginning as fast roll) which subsequently injects a slow-roll epoch motivates the scenario wherein inflation commences even as the inflaton is off its eventual attractor. This section is devoted to exploring some general implications of this type of dynamics for the CMB power spectrum.

We begin by considering scalar perturbations in terms of the Mukhanov--Sasaki (MS) variable ${\rm v}({\bf r},\eta)$, whose origin and applicability we briefly elaborate upon in Appendix~\ref{a1}. The starting point is provided again by the action \eqref{action_phi}, where the metric and the scalar field are both perturbed around some background solution $(\phi_0(\eta),a(\eta))$ as
\equ{pp2}{\phi({\bf r},\eta) \ = \ \phi_0(\eta) \, + \, \delta\phi({\bf r},\eta)\  }
and
\equ{pl2}{ds^2 \ = \ a^2(\eta)\bigg[\,-\,\big(1 \,+\, 2\psi({\bf r},\eta)\big)\, d\eta^2 \ +\  \big(1 \,- \,2\, \psi({\bf r},\eta)\big)\, d{\bf x} \cdot d{\bf x} \bigg] \ .}
Defining the MS variable
\equ{vdef2}{{\rm v}({\bf r},\eta) \, = \,  a(\eta)\left[\delta\phi({\bf r},\eta) \, + \, \frac{a(\eta)\, \phi_0^{\,\prime}(\eta)}{a^{\,\prime}(\eta)}\ \psi({\bf r},\eta)\right]\ ,}
where $a(\eta)$ is the conformal scale factor of a metric in the form \eqref{metric_conf} and ``primes'' denote derivatives with respect to the conformal time $\eta$, and letting
\equ{zdef2}{z(\eta) \, = \, a^2(\eta) \, \frac{\phi_0^{\,\prime}(\eta)}{a^{\,\prime}(\eta)} \ ,}
the quadratic action for ${\rm v}$ becomes
\equ{aso2}{S^{\,(2)} \ = \ \frac{1}{2}\, \int d^{\,3} {x}\ d \eta \left[({\rm v}^{\,\prime})^2 \ - \ \left(\nabla {\rm v}\right)^2 \ + \ \frac{z^{\,\prime\prime}(\eta)}{z(\eta)}\ ({\rm v})^2\right]\ .}
The advantage of working with the MS variable ${\rm v}$ is that all reference to the original background is confined to the time--dependent mass--like term
\be
W_S(\eta) \, = \, \frac{z^{\,\prime\prime}(\eta)}{z(\eta)} \ , \label{Ws}
\ee
where $z$ is defined in eq.~(\ref{zdef2}). A similar action, with $z(\eta)$ replaced by
\be
z_T(\eta) \ = \ a(\eta) \ = \ e^\frac{\cal A}{3} \ ,\label{zT}
\ee
can be used to describe tensor perturbations.

The MS variable is particularly convenient when comparing perturbations around two different homogeneous backgrounds that differ from each other perturbatively. For example, one can readily compare $n$-point correlation functions computed around the LM attractor of Section~\ref{sec:climbing}, for which \footnote{In the LM case, this expression applies for both scalar and tensor perturbations.}
\be
{W}^{(0)}(\eta) \, = \, \frac{1}{\eta^2}\ \frac{2\,-\,3\,\gamma^2}{(1\,-\,3\,\gamma^{\,2})^2} \ , \label{WLM}
\ee
with corresponding correlation functions computed around a different background with
\be
\frac{z^{\,\prime\prime}(\eta)}{z(\eta)} \ = \ {W}^{(0)}(\eta) \ + \ \delta W_S(\eta) \ .
\ee
In particular, if $\left|\frac{\delta W_S\,}{{W}^{(0)}}\right| \ll 1$ this can be done efficiently working in the interaction picture and treating $\delta W_S$ as a feeble interaction. The time evolution is determined by the operator
\equ{evop}{{\widehat U}\left(\eta,\eta_{\,0}\right) \ = \ T \, e^{\,-\,i\,\int^\eta_{\eta_0} d\eta\, {\widehat H}_{I}}\ }
via the interaction Hamiltonian
\equ{hintdef}{{\widehat H}_I\ = \ -\, \frac{1}{2}\, \int d^{\,3} k  \ \delta W_S(\eta) \, \left|{\widehat v}_{\bf k}(\eta)\right|^2\ ,}
with
\be
\widehat{v}({\bf r},\eta) \ = \ \int \frac{d^{\,3}k}{(2\pi)^\frac{3}{2}} \ \widehat{v}_{\bf k}(\eta) \ e^{i {\bf k} \cdot {\bf x}} \ ,
\ee
where ${\widehat v}_{\bf k}(\eta)$ is the spatial Fourier transform of the MS quantum field\footnote{In this paper quantum fields are denoted with a ``hat'' in order to distinguish from corresponding wavefunctions.} ${\rm \widehat v}({\bf r},\eta)$ and all operators evolve, as in the standard Heisenberg picture, with the unperturbed Hamiltonian determined by eq.~\eqref{aso2}. To first order, if $\eta_0$ indicates the time at which the initial state was defined, the effect on the connected n-point correlation function is
\equ{npmod}{\delta \langle \, {\widehat v}_{{\bf k}_1}(\eta)\, \ldots \, {\widehat v}_{{\bf k}_n}(\eta) \, \rangle_{\,in,\,in} \, = \, -\, i\, \int^\eta_{\eta_0}d\, \widetilde{\eta} \, \langle\, \left[{\widehat v}_{{\bf k}_1}(\eta)\, \ldots \, {\widehat v}_{{\bf k}_n}(\eta),\, H_I(\widetilde{\eta})\right]\, \rangle_{\,in,\,in} \ ,}
where the subscript on the expectation value indicates that, in the path--integral language, the expression ought to be evaluated resorting to the ``in--in'' Schwinger-Keldysh formalism \cite{schwingerkeldysh}. Combining (\ref{aso2}) with given vacuum initial conditions and applying Wick's theorem one can compute for instance the correction to the 2--point correlation function as
\begin{multline}
\delta \langle {\widehat v}_{{\bf k}_1}(\eta)\, {\widehat v}_{{\bf k}_2}(\eta) \rangle_{\,in,\, in} \ =  \\ i\, \delta^3({\bf k}_1 + {\bf k}_2)\int^\eta_{\eta_0}d \widetilde{\eta}~\delta W_S(\widetilde{\eta})\, \bigg[ G_{{k}_1}(\eta,\widetilde{\eta})\, G_{{k}_2}(\eta,\widetilde{\eta}) \, - \, G^\star_{{k}_1}(\eta,\widetilde{\eta})\, G^\star_{{k}_2}(\eta,\widetilde{\eta}) \bigg]\ , \label{npans}
\end{multline}
in terms of the Green functions for the LM attractor background of eq.~\eqref{metric_confLM},
\equ{gfsol}{G_{k}(\eta,\widetilde{\eta})\ = \ \frac{\pi}{4}\ \sqrt{\eta\,\widetilde{\eta}}\ H_\nu^{(1)}(-k\,\eta)\, H_\nu^{(2)}(-k\,\widetilde{\eta}) \quad\qquad (\eta > \widetilde{\eta}) \  ,}
where
\equ{nu}{\nu \ =  \ \frac{3}{2} \ \frac{1 \, - \, \gamma^{\, 2}}{1 \ - \ 3\, \gamma^{\, 2}} \ .}

The first--order correction to the 2-point correlation function picks up contributions from the integrand in eq.~(\ref{npans}) that end up superposing features and oscillations if $\delta W_S(\eta)$ has wide enough support in conformal time. It is instructive to see this in a concrete example, and to this end let us consider a step--wise perturbation. In order to simplify matters, let us also concentrate on the limiting case $\gamma=0$, which corresponds to standard de Sitter inflation, so that the corresponding Green function is simply
\be
G_{{k}}(\eta,\widetilde{\eta}) \ = \ \frac{1}{2\, k} \ \left(1 \, - \, \frac{i}{k\, \eta}\right)\left( 1 \, + \, \frac{i}{k\, \widetilde{\eta}} \right) \, e^{\,-\,ik(\eta-\widetilde{\eta})} \ . \label{greendesitter}
\ee
If a \emph{constant} $\delta W_S$ acts within the conformal time window $(-\eta_i,-\eta_f)$ with $\eta_{i,f}>0$, the integral in eq.~\eqref{npans} leads to an instructive deformation of the power spectrum, that as we discuss in the following section, can be defined in the present example through
\equ{deltaps2}{\delta^3 ({\bf k} + {\bf q})\ \delta P_{Q}(k) \ =\ k^3\, \lim_{\eta \, \to \, 0^-}  \ (\eta H)^2 \ \delta \, \langle \, 0|\, {\widehat v}_{\bf k}(\eta)\,
{\widehat v}_{\bf q}(\eta) \, |0 \rangle \ ,}
where
\be
\delta^3 ({\bf k} + {\bf q})\ P_{Q}(k) \ =\ k^3 \,\lim_{\eta \, \to \, 0^-} \langle \, 0|\, {\widehat {Q}}_{\bf k}(\eta)\,
{\widehat {Q}}_{\bf q}(\eta) \, |0 \rangle \ ,
\ee
and $\widehat{Q}({\bf r},\eta)$ is elaborated upon in Appendix \ref{a1}. Therefore:
\be
\delta P_{Q}(k) \ = \  k \,H^2\, \delta W_S\, \left[ \frac{1}{4}\ \cos(2k\eta_i)\, - \, \frac{\sin(2k\eta_i)}{2k\eta_i} \ - \ \frac{1}{4}\ \cos(2k\eta_f) \, + \, \frac{\sin(2k\eta_f)}{2k\eta_f}  \right] \ .  \label{constant_delta_desitter}
\ee
If the actual $W_S$ deviates from the attractor form of eq.~\eqref{WLM} for $-\eta_i < \eta < -\eta_f$, oscillations should show up in regions of the spectrum where $|k \eta_{i}|$ or $|k \eta_{f}|$ are of order one. In Section \ref{sec:numerics} we shall see this type of effects clearly in our numerical simulations. For the two--exponential system sizable oscillations draw their origin from early epochs close to the initial singularity, while additional features can be associated to a dip that, as we shall see in Section~\ref{sec:qma}, develops in $W_S(\eta)$ around the endpoint of the climbing phase. Its depth reflects the entity of the bounce against the hard exponential, and indeed the oscillations that it brings about are more pronounced for scalar fields that penetrate more deeply into the steep region $\phi>0$. Finally, whenever $|\eta_i|$ is very large or $|\eta_f|$ is very small the corresponding oscillations become essentially invisible, and indeed in Appendix~\ref{a3} we shall present a class of exact power spectra with a sizable infrared suppression but no oscillations.

Summarizing, one can anticipate the emergence of oscillatory features of scalar and tensor power spectra that typically present themselves even in the absence of initial particles. In the next sections we shall complement these arguments with numerical simulations and with a qualitative WKB analysis driven by analogies between the Mukhanov--Sasaki and Schr\"odinger problems. This last approach applies in principle even for large deviations from the attractor trajectory, and provides a simple rationale for the infrared suppression of scalar and tensor power spectra that is generic in models with an initial singularity. The overall IR modulation of the power spectra is also readily understood via spectral running arguments, which we discuss after presenting our results.

\scs{The power spectrum}\label{sec:powerspectrum}

In comparing the predictions of a particular model of the early universe to the observed CMB temperature anisotropies, one is required to compute the so--called dimensionless power spectrum, defined as
\equ{dps}{\delta^3 ({\bf k} + {\bf q}) P_Q(k) \ =\ k^3\,\langle \, 0| {\widehat Q}_{\bf k}(t)\, {\widehat Q}_{\bf q}(t) \, |0 \rangle \ .}
In this section and the following one we shall work in cosmological time, and we shall also find it convenient to work with the rescaled MS variable ${\widehat Q}_{\bf k}(t)$ of Appendix \ref{a1} that becomes independent of $t$ shortly after horizon crossing\footnote{Furthermore (see also Appendix~\ref{a1}), ${\widehat Q}_{\bf k}(t)$ remains well defined throughout the climbing phase, and at late times can be converted into the comoving curvature perturbations that source temperature anisotropies.}. ${\widehat Q}_{\bf k}(t)$ is a solution of the momentum--space counterpart of eq.~(\ref{qeom}),
\equ{qeomf}{{\ddot {\widehat Q}}_{\bf k} \ +\ 3\, H \, {\dot {\widehat Q}}_{\bf k} \ +\ \Bigl[\frac{k^2}{a^2} \ + \ V_{\phi\phi} \ - \ \frac{1}{a^3}\frac{d}{dt}\Bigl(\frac{a^3 \dot\phi_0^2}{H}\Bigr)\Bigr] {\widehat Q}_{\bf k}(t) \ =\ 0\ ,}
which describes perturbations around a background described by $a(t)$ and $\phi_0(t)$. As in previous sections, ``dots'' indicate derivatives with respect to the cosmological time $t$, and $V_{\phi\phi}$ is the second derivative of the potential with respect to $\phi$ evaluated on the background solution, that here is denoted by $\phi_0$.

The 2--point correlation function is usually computed in the vacuum state \emph{for the background solution}, for which ${\widehat a}_{\bf k}|0\rangle = 0$ for all ${\bf k}$, where ${\widehat a}^\dag_{\bf k}$ and ${\widehat a}_{\bf k}$ are creation and annihilation operators for the complete set of modes in terms of which the quantum field ${\widehat Q}(t,{\bf x})$ can be decomposed as
\be
\begin{split}
\label{qexp}{\widehat Q}(t,{\bf x}) \ &= \ \int \frac{d^{\,3}k}{(2\pi)^{3/2}} \ {\widehat Q}_{\bf k}(t)\, e^{i {\bf k} \cdot {\bf x}}\\
&= \ \int \frac{d^{\,3}k}{(2\pi)^{3/2}} \  \Bigl[u_{k}(t)\, e^{i {\bf k} \cdot {\bf x}}\, {\widehat a}_{\bf k} \ +\ u^\star_{k}(t)\, e^{-i {\bf k} \cdot {\bf x}}\, {\widehat a}^\dag_{\bf k}\Bigr]\ ,
\end{split}
\ee
where the $u_{k}$ are Klein-Gordon normalized solutions of (\ref{qeomf}). These modes will be different in general for different background trajectories $(\phi_0(t),a(t))$, but computing (\ref{dps}) in the vacuum state thus defined for our particular climbing trajectory yields the simple expression
\equ{dpsv}{P_Q(k) \ = \ k^3\, |u_{k}(t)|^2\ ,}
which is to be evaluated after horizon crossing.
\begin{figure}
\begin{center}
\epsfig{file=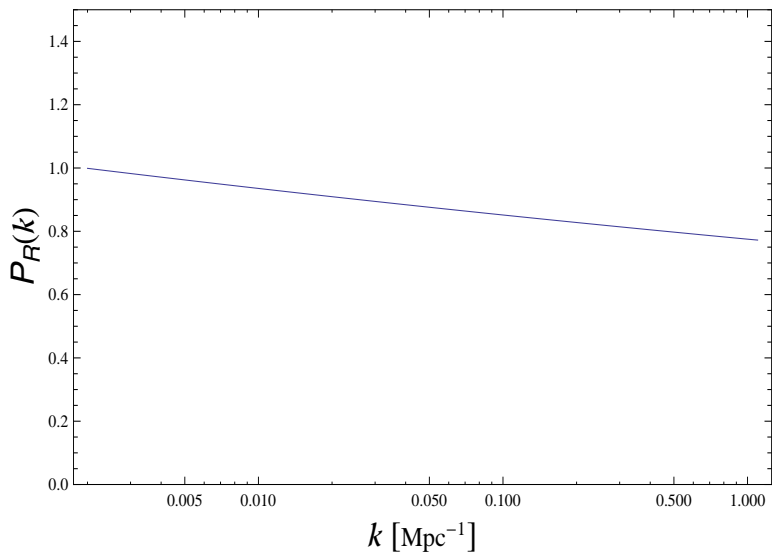, height=3in, width=4in}
\caption{Power spectrum for the LM attractor for $\gamma=\frac{1}{5\sqrt{6}}$, COBE normalized at $k = 0.002 {\rm Mpc}^{-1}$.}
\end{center}
\label{lmps}
\end{figure}
For a Universe that has always been on the LM attractor, the potential term within the square brackets in eq.~(\ref{qeomf}) vanishes and one is left with
\equ{qeomlm}{ {\ddot u}_{k} \ + \ \frac{1}{\gamma^{\,2}\, t}\ {\dot u}_{k} \ + \ \frac{k^2}{\left(\frac{t}{t_0}\right)^{\frac{2}{3\, \gamma^{\,2}}}}\
u_{k} \ =\ 0\ ,}
where we have chosen $t_0$ so that $a(t_0)=1$. The normalized modes corresponding to \eqref{qeomlm} are
\equ{qlms}{u_{k}(t) \ = \ \sqrt{\frac{3\, \gamma^{\,2}\,\pi \,t_0}{4(1\,-\,3\,\gamma^{\,2})}}\, \left(\frac{t}{t_0}\right)^{-\, \frac{1\,-\, \gamma^{\,2}}{2\, \gamma^{\,2}}}\ H^{(1)}_{\nu}\left(\frac{3\, \gamma^{\,2}\, k \,t \,{\left(\frac{t}{t_0}\right)^{-\,\frac{1}{3\, \gamma^{\,2}}}}}{1\,-\,3\,\gamma^{\,2}}\right) \ ,}
where $H^{(1)}_\nu(x)$ denotes a Hankel function of the first kind of order $\nu$, which is defined in eq.~\eqref{nu}. From this result and from the limiting behavior of the Hankel functions for small values of the argument \cite{A_S} one can see that
\be
k^3|u_{ k}(t)|^2 \ \sim \ k^{3-2\nu}
\ee
for long wavelengths, so that
the flat spectrum of de Sitter space is recovered in the limit $\gamma \to 0$. On the other hand, the case $\gamma=\frac{1}{5\sqrt{6}}$ corresponds to the spectrum displayed in fig.~1.

In evaluating the power spectrum for perturbations around a climbing solution, it is illuminating to allow for an initial state at $t_0$ that may not correspond exactly to the vacuum specific to the climbing phase. For example, the modes describing the vacuum for a background that is exactly on the LM attractor are certainly different, as is the case for the Bunch--Davies vacuum. Therefore, one may be more interested in computing
\equ{dpsg}{\delta^3 ({\bf k} + {\bf q}) P_{Q,\psi}(k) \ =\ k^3\,\langle \, \psi| {\widehat Q}_{\bf k}(t)\, {\widehat Q}_{\bf q}(t) \, |\psi\rangle_{in,in} \ ,}
where $\psi$ is an as yet unspecified state. For eigenstates of the number operator, or more generally tracing over a thermal density matrix, the result is a straightforward modification that reads
\equ{dprm}{P_{Q,N}(k) = k^3|u_{k}(t)|^2[1 + 2n(k)] \  ,}
where $n(k)$ is the relevant occupation number in the $k^{th}$ mode. This follows directly from the fact that in such states $\langle {\widehat a}_{\bf k} \rangle = \langle \widehat a^\dag_{\bf k} \rangle = 0$ while $\langle \widehat a_{\bf k} \widehat a^\dag_{\bf p}\rangle$ is $\delta$--function correlated\footnote{These are special cases of the so--called ``random phase'' states, for which these properties also hold \cite{kan1}.}. More generally one can have an initial state that fits into neither of these two categories, but can still be parametrized assigning to the modes corresponding to ${\bf k}$ occupation numbers $\beta_{\bf k}$. Specifically, on some fixed hypersurface, the initial data could correspond to all modes being in the vacuum of some background space--time other than what we are perturbing around (with associated mode functions $V_{k}(t)$). Such a state would appear to have finite occupation numbers with respect to the ${\widehat Q}(t,{\bf x})$ quanta (with mode functions $u_{k}(t)$). On this hypersurface, the two sets of modes can be related via Bogoliubov transformations of the form
\be
\begin{split}
\label{fexp}
V_{k} &= \cosh\theta_{k}\, u_{k} \ +\ e^{-i\delta_{k}}\, \sinh\theta_{k}\, u^\star_{k} \ , \\ V^\star_{k} &= \cosh\theta_{k}\,
u^\star_{k} \ + \ e^{i\delta_{k}}\, \sinh\theta_{k}\, u_{k} \ .
\end{split}
\ee
The initial state we are interested in is of the form $|\psi\rangle = {\widehat U}(\Theta) |0\rangle$, where ${\widehat U}$ is an operator that we shall
specify shortly. The correlation function can therefore be recast in the form $\langle \, 0 | {\widehat U}^{\dagger } {\widehat Q}_{\bf k}(t) {\widehat U} \,
{\widehat U}^{\dagger } {\widehat Q}_{\bf q}(t) {\widehat U} \, |0 \rangle_{in,in}$, so that it corresponds to performing a unitary transformation on the operator that has the effect of a Bogoliubov transformation on the states.

The quantum field producing particles of wavefunctions $V_k$ is
\begin{eqnarray}
{\widehat Q}' (t,{\bf x}) \ = \ \ {\widehat U}^{\dagger} \ {\widehat Q} (t,{\bf x}) \ {\widehat U} && =\ \int \frac{d^3k}{(2 \pi)^{3/2}} \ \Bigl[ V_{k}(t)\, e^{i {\bf k} \cdot {\bf x}}\, {\widehat a}_{\bf k} \ +\ V^\star_{k}(t)\, e^{-i {\bf k} \cdot {\bf x}}\, {\widehat a}^\dag_{\bf k}\Bigr] \ \\ \nonumber &&= \ \int \frac{ d^3k}{(2 \pi)^{3/2}} \ \Bigl[ u_{k}(t)\, e^{i {\bf k} \cdot {\bf x}}\, {\widehat b}_{\bf k} \ +\ u^\star_{k}(t)\, e^{-i {\bf k} \cdot {\bf x}}\, {\widehat b}^\dag_{\bf k}\Bigr] \ ,
\end{eqnarray}
where the two sets of creation and annihilation operators are related as
\be
\begin{split}
\label{car} {\widehat b}_{\bf k} &= \cosh\theta_{k}\, {\widehat a}_{\bf k} \ + \ e^{i\delta_{k}}\, \sinh\theta_{k}\, {\widehat a}^\dag_{\bf -k} \ , \\
{\widehat b}^\dag_{\bf k} &= \ e^{-i\delta_{k}}\, \sinh\theta_{k}\, {\widehat a}_{\bf -k} \ + \ \cosh\theta_{k}\, {\widehat a}^\dag_{\bf k} \ .
\end{split}
\ee
For transformations that only rotate modes up to some fixed momentum scale, the unitary transformation that formally effects this is\footnote{The Hilbert spaces that are related by the transformation in (\ref{fexp}) are unitarily equivalent so long as $\Theta_{k}$ has finite support, as will be the case in the examples of interest to us. If $\Theta_{k}$ had support for arbitrarily large values of $\bf k$ (\emph{e.g.} when $\Theta_{k}$ is a constant), the two Hilbert spaces would be unitarily inequivalent \cite{strocchi}.}
\be
{\widehat U} (\Theta) \ = \ e^{-\frac{1}{2}\, \int d^3k~[\Theta^\star_k \, {\widehat a}_{\bf k}  {\widehat a}_{\bf -k} \, -
\, \Theta_k\, {\widehat a}^{\dag }_{\bf k} {\widehat a}^{\dag }_{\bf -k} ]} \ , \label{UT}
\ee
with
\be
\Theta_{k} = \theta_{k} \, e^{i\delta_{k}} \ ,
\ee
and one can choose $\theta_k > 0$. The transformation $U$ is defined so that
\be
{\widehat U}^{\dagger}  \, {\widehat a}_{\bf k} \, {\widehat U}\ = \ {\widehat b}_{\bf k} \
\ee
and the result of starting with the initial data $V_k$ instead of $u_k$ is accounted for by the unitary transformation ${\widehat U}$ transforming the operator ${\widehat Q}$ in (\ref{qexp}) as ${\widehat Q} (t,{\bf x}) \ \rightarrow {\widehat Q}' (t,{\bf x})  =  {\widehat U}^{\dagger} {\widehat Q}
(t,{\bf x}) {\widehat U}$.

Relating the transformation (\ref{fexp}) to the standard Bogoliubov coefficients
\be
\alpha_{k} = \cosh\theta_{k} \ , \qquad \beta_{k} = e^{-i \delta_{k}}\sinh\theta_{k} \ ,
\ee
one can evaluate the 2-point correlation function in the state $|\psi\rangle = {\widehat U} (\Theta)|0\rangle$ as
\begin{eqnarray}
&& \langle \, \psi| {\widehat Q}_{\bf k}(t)\, {\widehat Q}_{\bf q}(t) \, |\psi\rangle_{in,in} \ \ = \
\langle \, 0| {\widehat Q}'_{\bf k}(t)\, {\widehat Q}'_{\bf q}(t) \,
|0 \rangle_{in,in}  =  \  |V_k|^2 \delta^3 ({\bf k} + {\bf q}) \nonumber \\
&&  = \, |u_{k}|^2 \left\{ 1 + 2 |\beta_{k}|^2 \ + \ 2 \ \cos (\delta_k + 2 \Delta_k) \ |\alpha_k \beta_k|  \right\}
\delta^3 ({\bf k} + {\bf q})    \ , \label{psi1}
\end{eqnarray}
where $u_{k} = e^{i \Delta_k}|u_{k}|$.
In position space, the power spectrum is encoded in
\be
\langle \, \psi| {\widehat Q}^2 (t,{\bf x}) \, |\psi\rangle_{in,in} \ = \ \int \frac{d^3 k}{(2 \pi)^3}\, |u_{k}|^2
\left\{ 1 + 2 |\beta_{k}|^2 \ + \ 2 \ \cos (\delta_k + 2 \Delta_k) \ |\alpha_k \beta_k|  \right\} \ , \label{psi2}
\ee
so that
\be
P_{Q,\psi}(k) = k^3|u_{k}(t)|^2\left\{ 1 + 2 |\beta_{k}|^2 \ + \ 2 \ \cos (\delta_k + 2 \Delta_k) \ |\alpha_k \beta_k|  \right\} \ . \label{psB}
\ee
As a result, the power spectrum picks up a modulation from the possible presence of initial particles. The effects of these initial excitations decay exponentially for all observable wavelengths, but their possible presence should be accounted for when considering CMB observables that were generated within a short window of inflation. In Appendix~\ref{a2} we show that for initial conditions that are natural for a Universe that emerged from a cold Big Bang dominated by the kinetic energy of moduli fields, the effects of any initial excitations are negligible. However, these effects may become significant if a large number of modes were in an initial state different from the one natural to the climbing phase.
\vskip 24pt


\scs{Numerics} \label{sec:numerics}
\begin{figure}[h!!]
\begin{center}$
\includegraphics[width=6.3in]{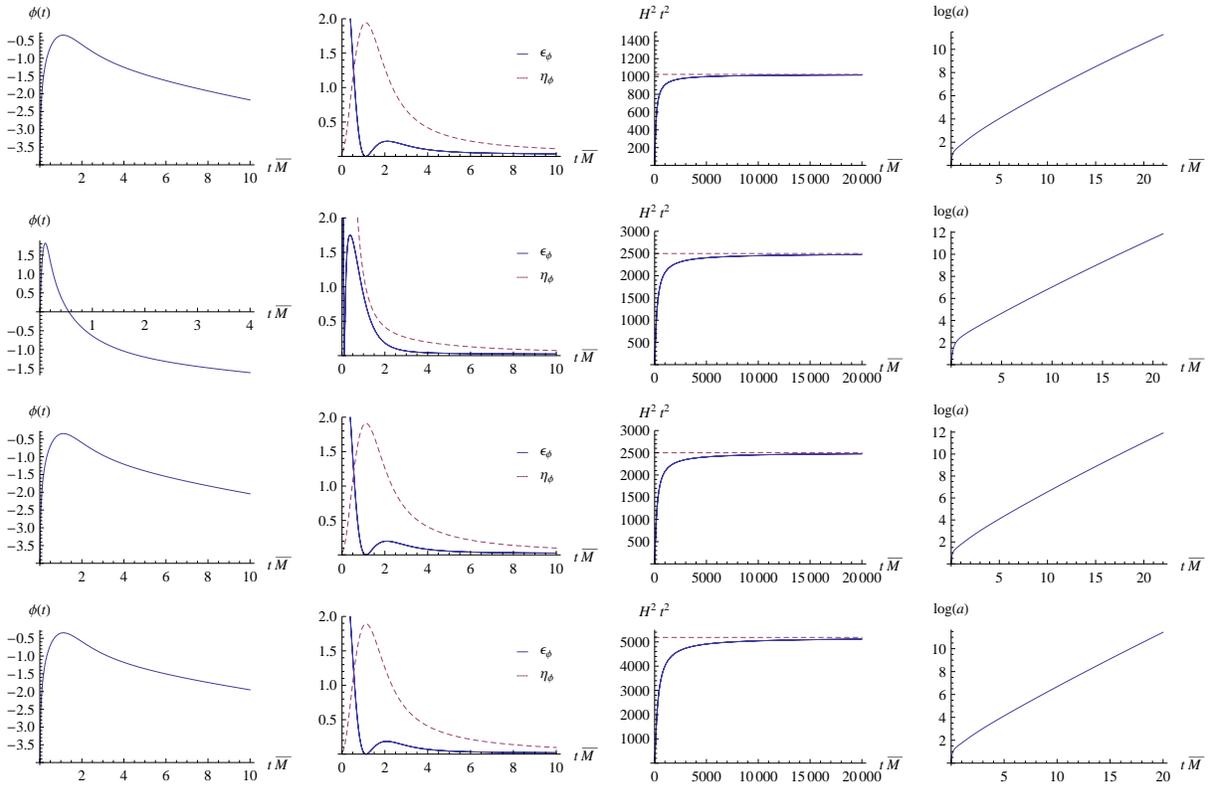}$
\end{center}
\caption{Background solutions for the potential \eqref{exppot2} with exponents $\gamma = \frac{1}{4\sqrt{6}}$ and $\phi_0 = -4$ (top), $\gamma = \frac{1}{5\sqrt{6}}$ (middle two, with $\phi_0 = 0,-4$ respectively) and $\gamma = \frac{1}{6\sqrt{6}}$, $\phi_0 = -4$ (bottom). The dashed line in the third column indicates the LM attractor for the respective single term potentials $V = \frac{M^2}{2}e^{\sqrt 6\gamma\phi}$.}
\label{bgsols3}
\end{figure}

In this section we present the results of our numerical integrations. They focus upon the simplest class of models involving a single scalar field minimally coupled to gravity that as we have seen in Section~\ref{sec:climbing}, combine an initial climbing phase with an eventual epoch of slow roll. The relevant potentials,
\be
V(\phi) \ = \ \frac{M^2}{2}\  \left( e^{\, \, \sqrt{6} \, \phi} \ + \ e^{\, \, \sqrt{6} \, \gamma\, \phi} \right) \ , \label{exppot1_r}
\ee
were given in eq.~\eqref{exppot2}, and we display them again here for the reader's convenience. They combine a critical exponential with another, milder exponential that is to sustain an eventual phase of slow--roll inflation. $\phi$ could represent the dilaton (or some other scalar field partly mixed with it), and therefore we shall be particularly interested in cases with a weak string coupling ($\phi < 0$), which are compatible with the climbing phenomenon. Starting with a particular value for $\phi_0$, which is kept as a free parameter, the initial speed for $\phi$ at very early times is determined by (\ref{climbing_phi}), to be expressed in cosmological time $t$, for any given slow--roll exponent $\gamma$. Letting\footnote{Where we remind again the reader that in restoring $\kappa$ in eq.~(\ref{exppot1_r}), the dimensionful pre--factor of the potential would read $\frac{M^2}{2\kappa^2}$, so that $\bar M$ has indeed the dimension of a mass once we set $\kappa =1$.} $\bar M^4 = \frac{M^2}{2}$, we integrated the background equation of motion for $\phi$ and the results are presented in fig.~\ref{bgsols3}. Our integrations begin at $t \, {\overline M} = 0.01$, where the scale factor is normalized to $a = 1$, with $a\to 0$ as $t\to 0$ representing the initial singularity.

Given that $H^2t^2 \to \frac{1}{9 \, \gamma^4}$ on the attractor, the third column of fig.~\ref{bgsols3} shows that the solutions approach the attractor asymptotically \emph{long after} the effective onset of inflation, as can be seen from the second column that displays the \emph{slow--roll parameters}\footnote{An alternative parametrization for the second slow roll parameter is sometimes taken to be $\bar \eta_\phi = -\frac{\ddot\phi}{H\dot\phi}$, which is related to $\eta_\phi$ above {\it during slow roll} as $\eta_\phi = \epsilon_\phi + \bar\eta_\phi$. We prefer the definition (\ref{sreta}), as this parameter is well defined throughout the range of our integrations.}
\begin{eqnarray}
\label{sreps}
\epsilon_\phi &\equiv& -\ \frac{\dot H}{H^2} \ ,\\
\label{sreta}
\eta_\phi &\equiv& \frac{V_{\phi\phi}}{V} \ ,
\end{eqnarray}
where $\phi_0$ is the background solution and $V_{\phi\phi}$ is the second derivative of $V$ with respect to $\phi$.
Note that the accelerated expansion ($\epsilon_\phi < 1$) begins within one ``string'' time $t \, {\overline M} \sim {1}$ in all examples, not much after the climbing phase has ended (left column of fig.~\ref{bgsols3}). Although $\eta_\phi$ takes slightly longer to converge to its attractor value, its relative size does not preclude the effective onset of accelerated expansion\footnote{We remind the reader that $\eta_\phi$ is only required to be small if its value is taken to be approximately constant during inflation. One can readily accommodate temporarily large values of $\eta_\phi$ consistent with persisting inflation provided it becomes negligible shortly thereafter, since the existence of an accelerated phase only requires that $\epsilon_\phi$ be small at any given time.}. We note in passing the brief spurt of de Sitter inflation ($\epsilon_\phi = 0$) around the inflection point. Although the solution closes in on the attractor only at rather late times $t \, {\overline M} \sim {\cal O}(10^4)$, inflation \emph{off the attractor} begins almost immediately. Although inflating off the attractor is far less effective (in the sense that far fewer $e$--folds of expansion occur in a given interval of time off the attractor than on it), a significant number of $e$--folds can still be attained within a relatively short time (as measured in units of ${\overline M}$).
\begin{figure}[h!!]
\begin{center}
\includegraphics[width=6in]{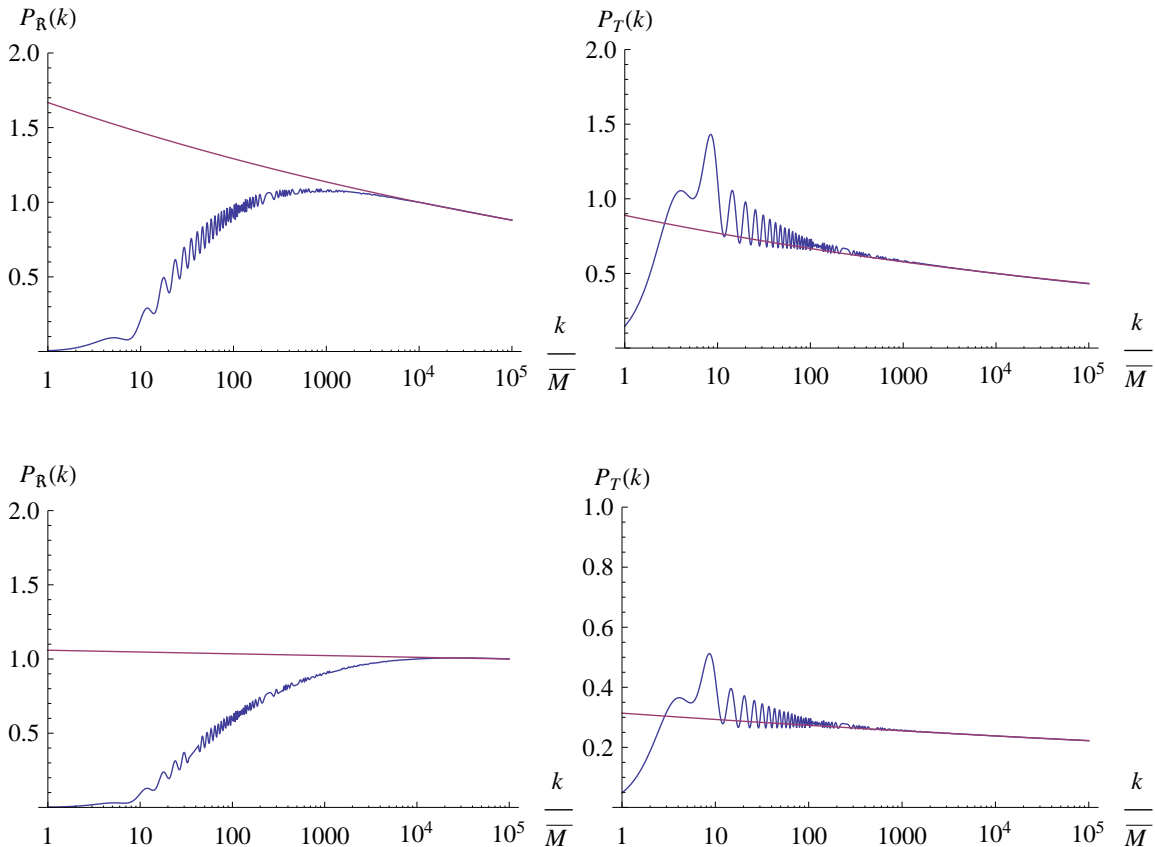}
\end{center}
\caption{Scalar (left) and tensor (right) spectra for $\gamma = \frac{1}{4\sqrt{6}}, \phi_0=-4$ (top), and $\gamma = \frac{1}{6\sqrt{6}}, \phi_0 = -4$ (bottom)-- best fit at short wavelengths with attractor spectra for $n_S=0.944, n_T=0.9375$ and $n_S=0.995, n_T=0.97$, respectively. The comoving wavenumber $k$ is in units of ${\overline M}$. The LM attractor would have led to $n_S = n_T = 0.9375$ for $\gamma = \frac{1}{4\sqrt{6}}$ and $n_S= n_T = 0.972$ for $\gamma = \frac{1}{6\sqrt{6}}$ .}
\label{spectra}
\end{figure}
\begin{figure}[h!]
\begin{center}
\includegraphics[width=6in]{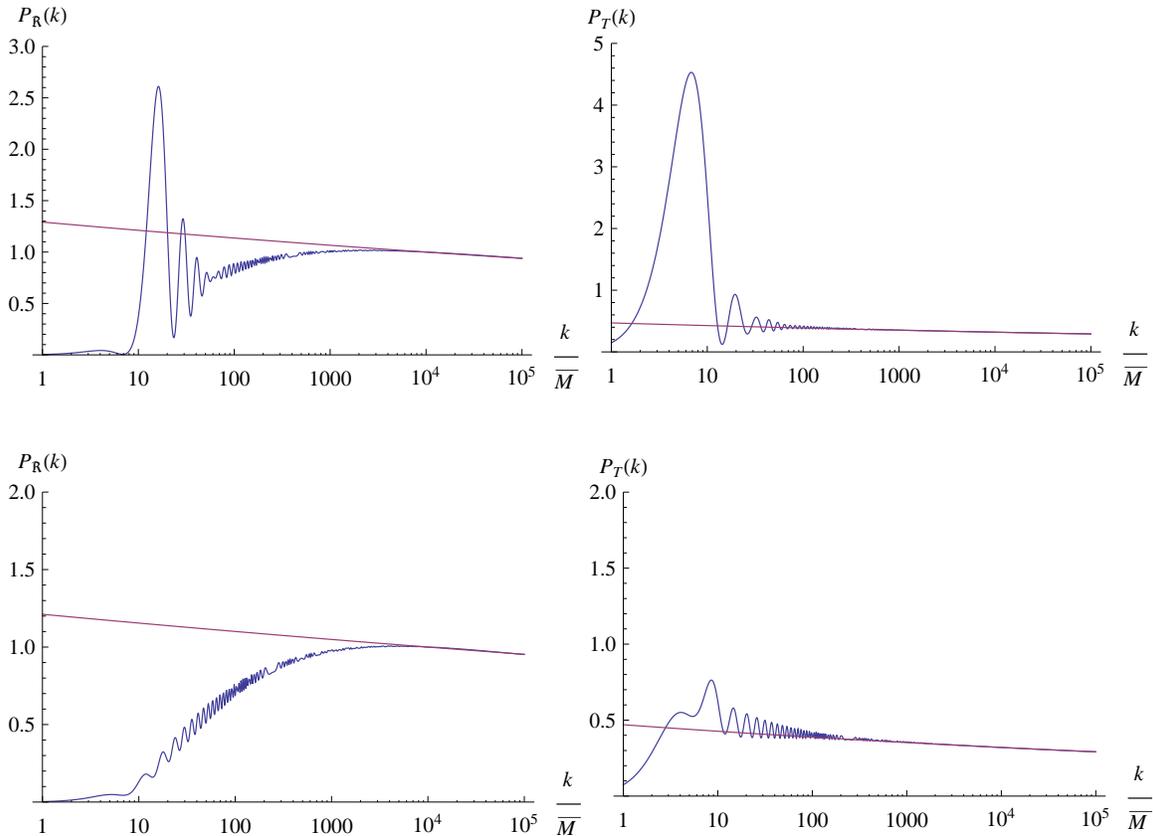}
\end{center}
\caption{Scalar (left) and tensor (right) spectra for $\gamma = \frac{1}{5\sqrt{6}}$, with $\phi_0 = 0$ (top), and $\phi_0 = -4$ (bottom) -- best fit with attractor spectra for $n_S=0.979, n_T=0.958$ and $n_S=0.975, n_T=0.958$ respectively. The comoving wavenumber $k$ is in units of ${\overline M}$. The LM attractor would have led to $n_S = n_T = 0.96$.}
\label{spectra10}
\end{figure}

We have computed the power spectra for a range of slow--roll exponents in figs.~\ref{spectra} and \ref{spectra10}. We note that, in addition to superposed oscillatory features, \emph{at long wavelengths there is a marked suppression of power in the scalar spectrum and a corresponding, if less pronounced, enhancement of power in the tensor spectrum}. We have adapted these results to the best fit spectra generated by equivalent solutions that are perpetually on the attractor, determined by the criterion that for the shorter wavelengths (spanning the majority of the scales that we can access with current observations) the two types of spectra coincide \footnote{Short wavelengths in this case are defined demanding that the climbing phase be sufficiently outside our present horizon and that at some convenient pivot scale (e.g. $0.0805 {\rm Mpc}^{-1}$) as defined by the horizon, the two spectra be COBE normalized. For the purpose of our plots, we have taken this scale to define the comoving wavelength $k = 10^4\bar M$.}.

In each case, the plots represent comoving wavelengths that have exited the horizon over approximately 12 $e$--folds of inflation. Given that the range we access in WMAP data constitutes a mere window of 6 $e$--folds, we see that we are at most privy to a subset of the modes represented in these plots. Although we would have to be exceedingly privileged to access wavelengths that exited right as inflation began\footnote{As we have stressed in the Introduction, however, this possibility is to be rightfully considered \cite{just} in the event that it may explain some large--scale features that do not fit comfortably within $\Lambda$CDM models.}, we see that one can still expect to find some trace of the climbing phase if the largest modes that are accessible to us left the horizon no later than 6-7 $e$--folds after the climbing phase. This is due to the fact that although inflation effectively started right away (within one string time $t \, {\overline M} \sim 1$ of the end of the climbing phase), the eventual approach to the attractor takes far longer. This is illustrated by the third column of fig.~\ref{bgsols3}.

We have also evaluated the multipole moments of the angular power spectrum for the climbing solutions corresponding to $\gamma = \frac{1}{4\sqrt{6}}$ and $\frac{1}{6\sqrt{6}}$ respectively, and we have compared them to the corresponding best--fit power--law spectra of figs.~\ref{spectra} and \ref{spectra10}. For this, we simply used the fact that at the largest angular scales ($\ell \leq 30$), to an excellent approximation \cite{cmbslow}
\equ{clan}{C_\ell \ = \ \frac{2}{9\pi}\ \int \frac{dk}{k}\ \mathcal P_\mathcal R(k)\ j^{\,2}_\ell[k(\eta_0-\eta_r)]\ ,}
where $j_\ell$ is a spherical Bessel function and $\eta_0-\eta_r$ denotes our present comoving distance to last scattering surface. In this fashion we find again that if inflation had started within 6-7 $e$--folds of our present horizon exit, the climbing phenomenon would bring about a noticeable drop in power at the largest angular scales that would become more significant the closer the climbing phase were to the exit of our current horizon.

\begin{figure}[h!!]
\begin{center}$
\begin{array}{cc}
\includegraphics[width=3in]{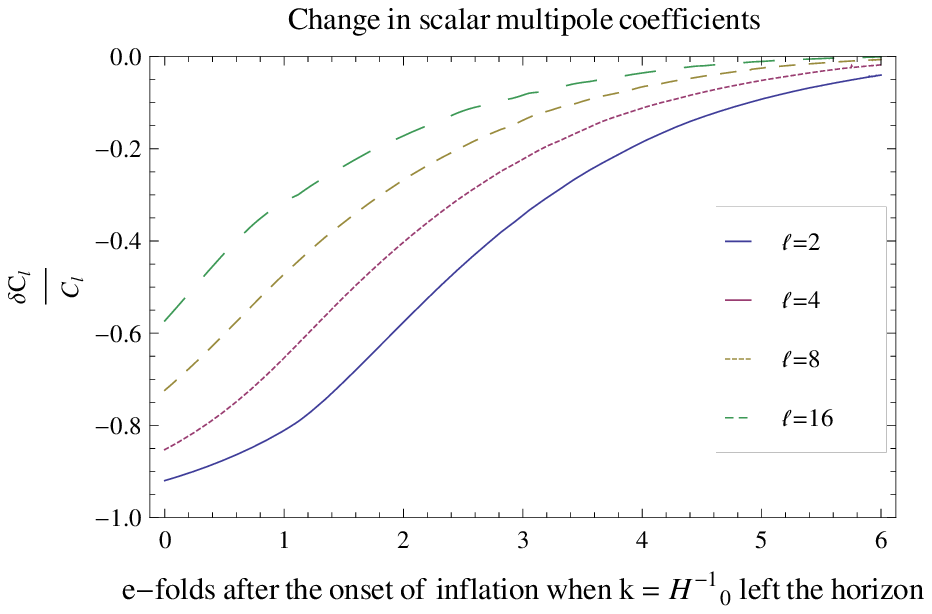} &
\includegraphics[width=3in]{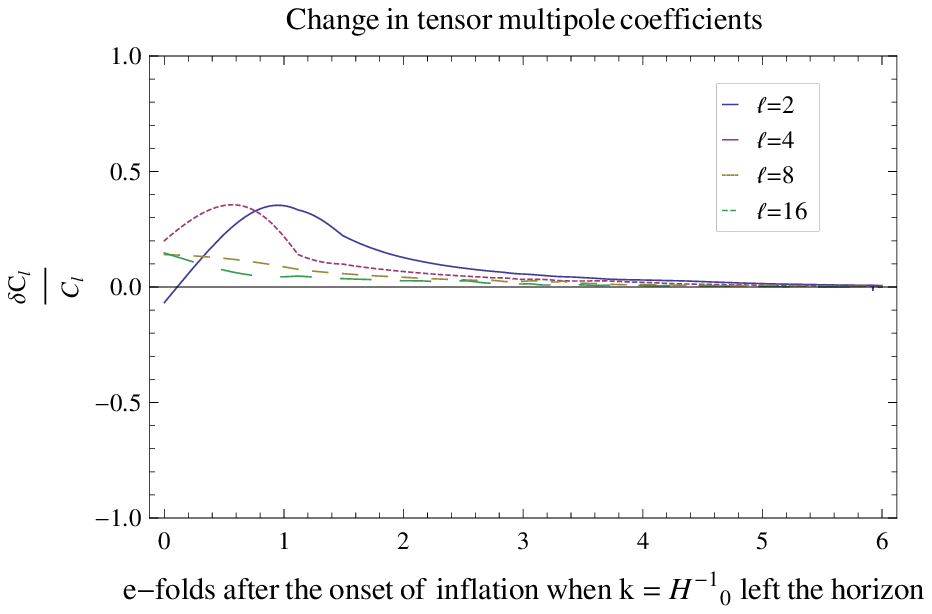}\\
\includegraphics[width=3in]{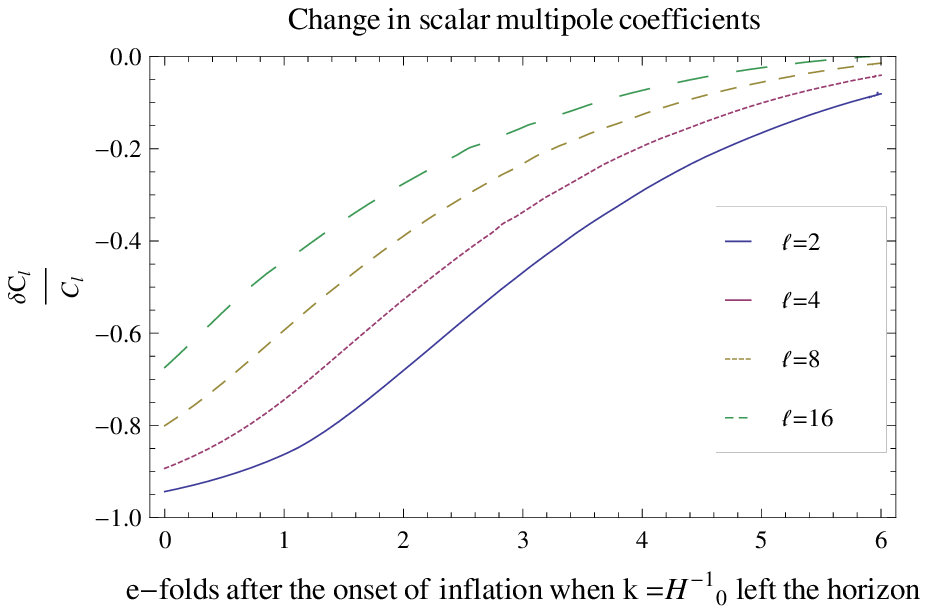} &
\includegraphics[width=3in]{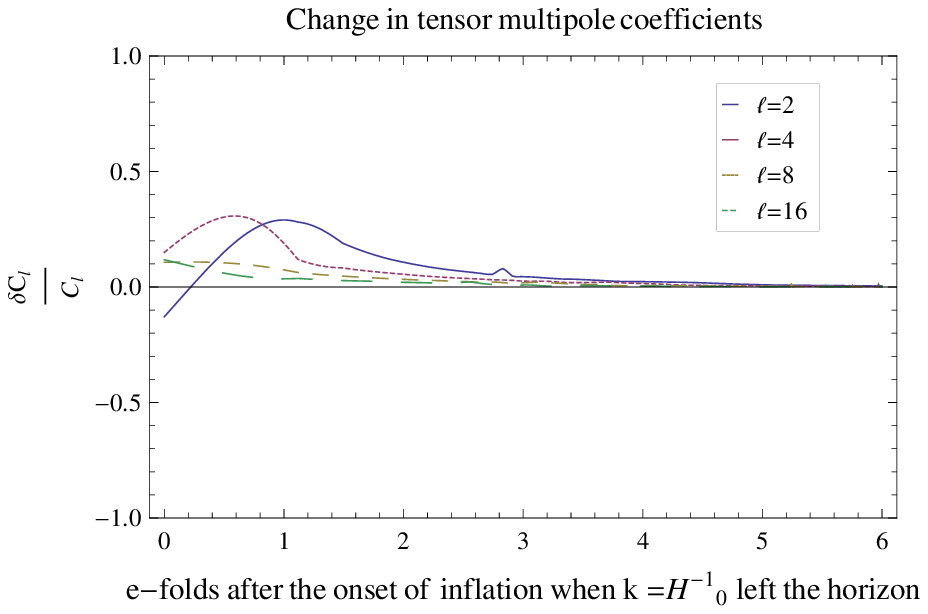}
\end{array}$
\end{center}
\caption{Changes in the scalar and tensor multipole coefficients for $\gamma = \frac{1}{4\sqrt{6}}$ with respect to featureless spectra with $n_S= 0.944, n_T= 0.9375$ (top), and for $\gamma = \frac{1}{6\sqrt{6}}$ with respect to featureless spectra with $n_S= 0.995, n_T=0.97$ (bottom).}
\label{cl}
\end{figure}

We note that the drop in power for a range of parameters (depending on the multipole in question) becomes \emph{more significant than the uncertainty implied by cosmic variance},
\equ{cv}{\left|\frac{\Delta C_\ell}{C_\ell}\right| \ = \ \sqrt{\frac{2}{2\ell + 1}} \ , }
for each of the multipoles represented in fig.~\ref{cl}, if $k = H_0^{-1}$ had exited the horizon during inflation no more than $2$ $e$--folds after inflation began. Therefore a statistically significant drop in the power at long wavelengths can result if the climbing phase occurred within $3.5$ $e$--folds\footnote{In all examples that we have displayed, with $\gamma=\frac{1}{4\sqrt{6}}$ and $\frac{1}{6\sqrt{6}}$, about $1.5$ $e$--folds of expansion resulted before slow-roll was effectively attained.} of the horizon exit for the largest scales that are presently accessible to us. It is tempting to infer that this may be at least partly responsible for the apparent large--angle suppression of power in the CMB \cite{low_power}.

\begin{figure}[h!]
\begin{center}$
\includegraphics[width=4.5in]{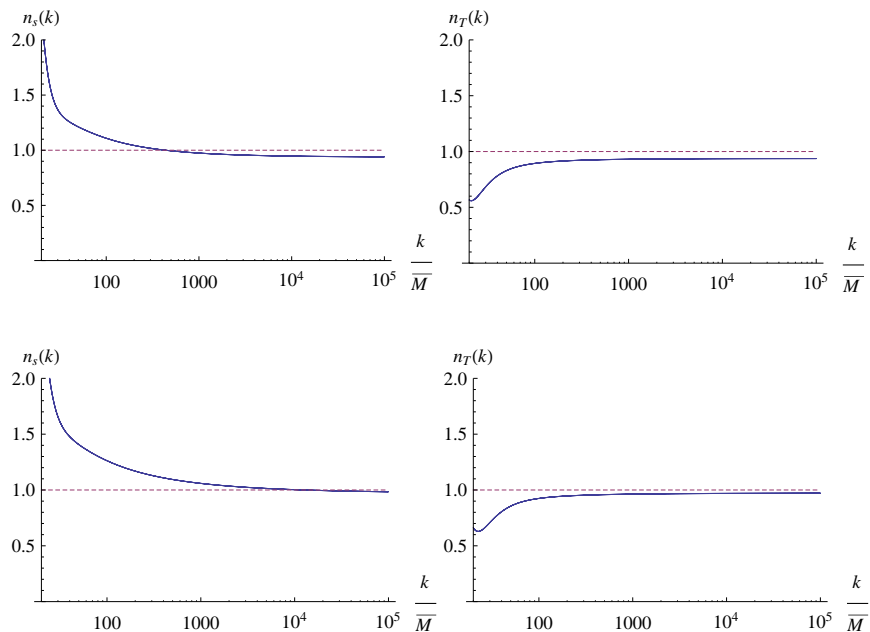}$
\end{center}
\caption{Running spectral indices for $\gamma = \frac{1}{4\sqrt{6}}$ (top) and $\gamma = \frac{1}{6\sqrt{6}}$ (bottom). The comoving wavenumber $k$ is in units of ${\overline M}$.}

\label{run}
\end{figure}

It is informative to compare the best--fit values of $n_S$ and $n_T$ obtained in figs.~\ref{spectra} and \ref{spectra10} (indicated in the respective captions) to those of corresponding solutions that would have always been on the LM attractor corresponding to the particular exponent $\gamma$. Up to the expected higher--order corrections coming from the time variation of $\epsilon_\phi$ and $\eta_\phi$, they are in reasonable agreement at short wavelengths. Furthermore, one can readily understand the gross features of scalar and tensor spectra starting from the spectral relations for solutions that \emph{remain on the attractor} \cite{lyth_liddle}:
\begin{eqnarray}
\label{tilts}
n_S \ - \ 1 &=& 2(\, \eta_\phi \ - \ 3\, \epsilon_\phi) \ ,\\
\label{tiltT}
n_T \ - \ 1 &=& -\ 2\, \epsilon_\phi \ .
\end{eqnarray}
For the single--exponential potential \eqref{exppot1}, \emph{on the attractor}, $\epsilon_\phi = 3\,\gamma^2$, $\eta_\phi = 6\,\gamma^2$ and so $n_S-1$ $=n_T-1$ $= -\, 6\, \gamma^2$. Reconsidering now eqs.~(\ref{tilts}) and \eqref{tiltT} while allowing for spectral running (\emph{i.e.} allowing for $n_S = n_S(k), n_T = n_T(k)$), and deriving these from the values obtained for $\epsilon_\phi(t_k)$ and $\eta_\phi(t_k)$, evaluated at the time at which comoving wavenumber $k$ exits the horizon\footnote{Given the fact that $\dot\epsilon_\phi = \dot\eta_\phi = \mathcal O(\epsilon_\phi^2,\eta_\phi^2,\eta_\phi\epsilon_\phi)$, allowing for such a spectral running will be accurate up to errors quadratic in the slow--roll parameters.}, one can arrive at the running spectral indices plotted in fig.~\ref{run}.
We note that in each of the plots, the spectral indices run towards their attractor values for short wavelengths (as the solution settles towards the LM attractor), but deviate considerably for longer wavelengths. We can now readily understand the long wavelength modifications of scalar and tensor spectra -- from (\ref{tiltT}) we see that as $\epsilon_\phi$ monotonically decreases toward its asymptotic attractor value as it emerges from the climbing phase, $n_T-1$ approaches $-6\, \gamma^2$ from below. This means that for longer wavelengths, the spectrum had an even redder tilt, hence accounting for the convexity of the envelope for the modulations of the tensor spectra in figs.~\ref{spectra} and \ref{spectra10}, consistently with the enhancement of power at the longest wavelengths. For the scalar spectra, from eq.~(\ref{tilts}) we find that the slower convergence of $\eta_\phi$ to its attractor value compared to $\epsilon_\phi$ results in a positive (blue) spectral tilt for longer wavelengths, which eventually asymptotes to its negative (red) attractor value\footnote{Note the coincidence of the inflection points of the scalar spectra in fig.~\ref{spectra} with the transition of the spectral index from blue to red in fig.~\ref{run}.}. This accounts for the concavity of the low--frequency portion of the scalar power spectrum, and hence for the wide suppression of scalar power at long wavelengths.

One can also understand the qualitative effects of starting inflation off the attractor, but now initially descending, which would be possible if the first exponential in eq. \eqref{exppot1_r} were replaced by a slightly milder one. In this case the spectral index $n_S(k)$ would start out again larger than one, but cosmological friction would make it converge to one more rapidly, resulting in a less relevant suppression of the scalar spectrum that would be confined to much longer wavelengths in comparison with the initially climbing solution. All in all, the effects of pre--inflationary dynamics are far more significant for a solution that initially climbs, since the scalar field comes close to the attractor trajectory much later in this case.

Although we find a noticeable suppression for the lowest multipoles if the largest scale we observe today exited the horizon within 6-7 e--folds of the climbing phase, this drop would be \emph{statistically significant} \footnote{In the sense that the drop in power for the low multipoles would surpass the threshold obscured by cosmic variance.} only if horizon exit occurred within 3-4 e--folds. However, the consistency of the suppression over a wide range of multipoles ought to boost its significance even if horizon exit occurred at later times. Furthermore, if for various parameters of the potential and initial conditions for the inflaton, the superposed oscillations persisted to shorter wavelengths to a significant enough degree, the combination of the two effects might be easier to discriminate and constrain in the data. We leave to a future study a full analysis of any improved fit to WMAP data that may be offered by a climbing phase. However, in merely setting out to understand the generic consequences of a class of pre--inflationary dynamics inherent in an interesting class of string constructions, we find it encouraging that large--angle anomalies in the CMB \cite{low_power} may find a ready and natural explanation.
\vskip 24pt

\scs{The Schr\"odinger analogy}\label{sec:qma}

Further qualitative insight into the modifications of the CMB power spectrum can be attained in general via WKB methods. The MS equation that results from (\ref{aso2}) has in fact the Schr\"odinger--like form
\be
\left( \frac{d^2}{d\eta^2} \ + \ p^2(k,\eta) \right) v_{k}(\eta) \, = \, 0 \ , \label{mseta}
\ee
where
\be
p^2(k,\eta) \, = \, k^2\ -\ W_S(\eta)
\ee
and where the MS ``potential'' $W_S$ is defined in eq.~\eqref{Ws}~\footnote{As we have seen, the corresponding MS equation for tensor perturbation is identical, up to the replacement of $W_S$ with the ``potential'' $W_T$ determined by eq.~\eqref{zT}.}.

\begin{figure}
\begin{center}
  \epsfig{file=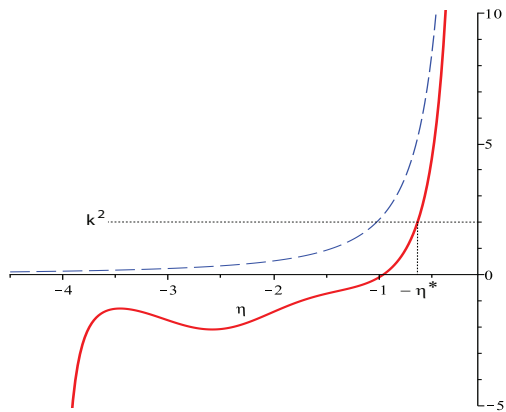, height=2.2in, width=2.75in}
  \caption{$W_S(\eta)$ for a climbing solution in the potential \eqref{exppot2} with $\gamma=0.1$ (continuous red), and ${W}^{(0)}(\eta)$ for the corresponding LM attractor in the potential \eqref{exppot1} (dashed blue).}  \label{lmwrfig}
\end{center}
\end{figure}

Notice that some key information is generally available on $W_S$ for this type of systems, which determines completely its asymptotic behavior at the initial singularity and at late times. One can indeed show, starting from the results in Section~\ref{sec:climbing}, that
\be
W_S \ \ {\phantom a}_{\widetilde{\eta \to - \eta_0}} \ \ - \ \frac{1}{4} \, \frac{1}{(\eta+\eta_0)^2} \ ,\label{WsBB}
\ee
where $- \eta_0$ (with $\eta_0 > 0$) corresponds to the initial singularity and is determined by the condition that the conformal time $\eta$ tend to zero for large values of $t$ or $\tau$. In addition the systems of interest, with $0<\gamma<\frac{1}{\sqrt{3}}$, approach an eventual slow--roll phase, so that
\be
W_S \ \ {\phantom a}_{\widetilde{\eta \to 0}} \ \ \frac{\nu^2\,-\, \frac{1}{4}}{\eta^2} \ ,\label{Wsfinal}
\ee
where $\nu$ is defined in eq.~\eqref{nu}, since $W_S$ is to attain the limiting LM behavior \eqref{WLM}.
These two limiting behaviors make it inevitable for $W_S$ to cross the real axis for a certain value of the conformal time in the interval $(-\eta_0,0)$, as can be seen in fig.~\ref{lmwrfig}. As we shall see, this generic structure suffices to bring along a generic depression of the power spectrum of CMB fluctuations in the infra--red in models with an initial singularity.

There is therefore a striking analogy with one--dimensional Quantum Mechanics, but for an important proviso. Systems describing an eventual inflationary phase map into an \emph{initial--value} Schr\"odinger--like problem featuring an impenetrable barrier that, as such, overcomes any available ``energy'' $k^{\,2}$ for small negative values of $\eta$, as can be seen in fig.~\ref{lmwrfig}. In ordinary Quantum Mechanics, a barrier of this sort would result in total reflection, and thus in a stationary wave in the classically allowed region, accompanied by a decaying mode in the classically forbidden region. In the Mukhanov--Sasaki problem, however, in the intermediate region where $W_S(\eta)$ is small, $v_{k}(\eta)$ and its complex conjugate ought to be close to incoming and outgoing plane waves if close to the initial singularity the system is to lie in a Bunch--Davies--like vacuum. In this fashion, beyond the barrier these types of solutions leave way in general to combinations of a decaying and a growing mode, the latter of which readily dominates. In a cosmological setting, this very behavior reflects itself in the growth of perturbations after horizon crossing. The gross features of the phenomenon are well captured by the conventional WKB matching relations, which in the classically forbidden region lead to
\be
v_{k}(\,-\,\epsilon)\ \sim \ \frac{1}{\sqrt{2\, |p(-\epsilon)|}} \ \exp\left(\int_{-\eta^\star}^{-\epsilon} |p(y)| \, dy \right)  \ , \label{WKBlim}
\ee
where the small positive number $\epsilon <<1$ identifies the end of inflation, and where at early enough initial times an oscillatory behavior like
\be
v_{k}(\eta)\ \sim \ \frac{1}{\sqrt{2\, k}} \ e^{\,-\, i\, k\, \eta} \
\ee
would correspond to the Bunch--Davies vacuum of de Sitter inflation. In eq.~\eqref{WKBlim} $-\eta^\star$ is the classical turning point of fig.~\ref{lmwrfig}, such that $W_S(-\eta^\star)=k^2$, and one is interested in the leading behavior for $\epsilon \to 0$, after several $e$--folds of inflation.

For a scalar field on the LM attractor corresponding to the milder exponential potential of the previous section,
\be
V \, = \,  \frac{M^2}{2} \ e^{\, 2\, \gamma\, \vf}
\ee
with $\gamma < \frac{1}{\sqrt{3}}$, eq.~\eqref{mseta} reduces to
\be
\left[ \frac{d^2}{d\eta^2} \ + \ \left( k^2\ -\ \frac{\nu^{\,2} \, -\, \frac{1}{4}}{\eta^{\, 2}}  \right) \right] v_{k}(\eta) \, = \, 0 \ , \label{MSattractor}
\ee
where $\nu$ is defined in eq.~\eqref{nu}.
One can appreciate the nature of the leading WKB approximation comparing it, for the LM attractor, to the exact solutions of eq.~\eqref{MSattractor}. As is well known, these can be expressed in terms of Hankel functions as
\be
v_{k}(\eta) \, = \, e^{\, i \, \frac{\pi}{2} \left( \nu \, +\, \frac{1}{2} \right)}\ \sqrt{\frac{(-\eta)\pi}{4}} \ H_{\nu}^{(1)}(-k \eta) \ , \label{Slm}
\ee
so that the limiting behavior for $\eta \to -\epsilon$ \cite{A_S},
\be
v_{k}(\eta) \, \sim \, \frac{1}{\sqrt{2\, \pi}} \ e^{\, i \, \frac{\pi}{2} \left( \nu \, -\, \frac{1}{2} \right)}\ \Gamma(\nu) \ k^{\, -\, \nu} \, \left(\frac{\epsilon}{2} \right)^{\frac{1}{2} \, - \, \nu}  \ ,
\ee
results in the power spectrum
\be
P_{\cal R} (k) \ \sim \ {k^{\,3}} \ \left| \frac{v_{k}}{z} \right|^2 \ \sim \ k^{\,3 \, - \, 2 \, \nu} \ , \label{speclm}
\ee
with the characteristic $3 - 2 \nu$ exponent.

The essential features of this expression are qualitatively, albeit not exactly, captured by the leading WKB formula \eqref{WKBlim}, which up to overall phase factors yields
\be
v_{k}(\eta) \, \sim \, \frac{1}{\sqrt{2}} \ \epsilon^{\,\frac{1}{2}\,- \, {\widetilde{\nu}}}\ \left(\frac{2}{k}\right)^{{\widetilde{\nu}}}\ \left({\widetilde{\nu}}\right)^{\, {\widetilde{\nu}} \, -\, \frac{1}{2}}\ e^{\,-\, {\widetilde{\nu}}} \ , \label{Swkblm}
\ee
where
\be
{\widetilde{\nu}} \, = \, \sqrt{\nu^{\, 2} \, -\, \frac{1}{4}} \ .
\ee
The WKB amplification factor of eq.~\eqref{Swkblm} thus recovers the qualitative features of the power spectrum of eq.~\eqref{speclm}, but getting the precise slope would entail the replacement of $\nu$ with ${\widetilde{\nu}}$, taking into account higher--order corrections~\footnote{The apparent discrepancy indeed disappears if one takes into account higher--order WKB corrections, which recover the expansion of $\nu$, expressed as $\widetilde{\nu}\sqrt{1+\frac{1}{4\,\widetilde{\nu}^2}}$, in inverse powers of $\widetilde{\nu}$, as the reader can verify for the next few orders. The successive orders of the WKB expansion are in fact sized by the relative variation of the potential within spatial distances comparable to a wavelength, which is smaller for larger values of $\nu$ or $\widetilde{\nu}$.}.

Together with the definition \eqref{speclm} of the power spectrum, eq.~\eqref{WKBlim} readily accounts for a generic suppression of the low--frequency tails of scalar and tensor power spectra in models with an initial singularity. For any given $k$, if one begins in the generalization of the Bunch--Davies vacuum for negative $\eta$'s in the wide flat region of $W_S(\eta)$, the amplification of $v_{k}(\eta)$ at horizon crossing reflects the area below $|p(\eta)|$, or equivalently the contribution of regions of conformal time where $W_S(\eta) \geq k^2$ (see fig.~\ref{lmwrfig}). As we have stressed, $W_S$ is bound to be negative at early times, so that the amplification factor cannot grow beyond limit for small values of $k$, and precisely in this region the overall $k^3$ of $P_{\cal R}(k)$ takes over.

On the other hand, the WKB formula \eqref{WKBlim} does not capture an interesting feature related to the dip present in fig.~\ref{lmwrfig}, which belongs to a region where $W_S$ is negative and is roughly centered around the conformal time when the scalar inverts its motion. One is thus led to expect that oscillations be present in the power spectrum, as for the transmission past an attractive well in Quantum Mechanics. This reasoning is along the lines, and actually equivalent to, the perturbative argument presented in Section~\ref{sec:perturbations}. As we saw in Section~\ref{sec:numerics}, oscillations are indeed generically present in the power spectra.
\begin{figure}
\begin{center}
  \epsfig{file=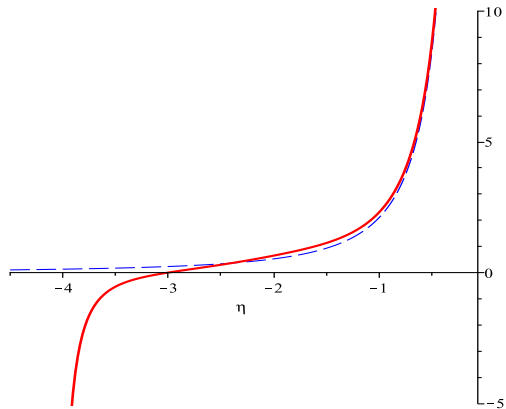, height=2.2in, width=2.75in}
  \caption{$W_T(\eta)$ for a climbing solution in the potential \eqref{exppot2} with $\gamma=0.1$ (continuous red), and ${W}^{(0)}(\eta)$ for the corresponding LM attractor in the potential \eqref{exppot1} (dashed blue).}  \label{lmwrfig_T}
\end{center}
\end{figure}

Fig.~\ref{lmwrfig} displays typical forms of $W_S(\eta)$ for the two cases of interest, an LM attractor and the climbing solution for the two--exponential potential of eq.~\eqref{exppot2} that eventually approaches it. The climbing solution emerges from an initial singularity at a finite negative value of $\eta$, where $W_S(\eta)$ has a vertical asymptote, to then approach for $\eta \to 0$ the curve ${W}^{(0)}(\eta)$ of eq.~\eqref{WLM}. Actually, $W_S(\eta)$ for the climbing scalar lies always below ${W}^{(0)}(\eta)$ and comes close to it rather late, so that for relatively small values of $k$ the integral in eq.~\eqref{WKBlim} is reduced with respect to its attractor value for two conspiring reasons. First, a lower $W_S(\eta)$ for the climbing scalar implies that the lower end $-\eta^\star$ of the integral in \eqref{WKBlim} lies to the right of the corresponding value for the attractor, and in addition the area below the curve, and thus below $|p(\eta)|$, is also reduced. All in all, one can thus see by inspection that the WKB amplification factor in eq.~\eqref{WKBlim} is reduced for the climbing scalar, and more so for smaller values of $k$. On the other hand, for the small negative values of $\eta$ that are relevant for large values of $k$, $W_S(\eta)$ approaches from below ${W}^{(0)}(\eta)$, and therefore the power spectra of climbing scalar and attractor solutions ought to merge in the ultraviolet region. Conversely fig.~\ref{lmwrfig_T} indicates that, tracing it backwards from the high--frequency region where it falls on the LM form of eq.\eqref{speclm}, the power spectrum of tensor perturbation ought to be initially \emph{slightly enhanced} in the infrared region with respect to the attractor solution, before being finally suppressed at very low frequencies as $W_T(\eta)$ crosses the real axis. This is all along the lines of Section \ref{sec:numerics}, and actually the relative behavior of scalar and tensor perturbations can be foreseen rather clearly via the WKB method, comparing the two expressions
\be
W_S \, \equiv \, \frac{z^{\,\prime\prime}}{z}\, = \, 3 \ e^{\,\frac{2}{3}\, {\cal A}} \left[ - \ \frac{1}{2}\ V_{\vf\vf} \ - \ 2\,  \frac{d \vf}{d\tau} \, \frac{V_{\vf}}{\sqrt{1\,+\, \left(\frac{d \vf}{d\tau}\right)^2}}\ +\ \frac{V}{9} \ \frac{2\, - \, 17\, \left(\frac{d \vf}{d\tau}\right)^2 \, -\, \left(\frac{d \vf}{d\tau}\right)^4}{1\,+\, \left(\frac{d \vf}{d\tau}\right)^2} \right] \label{Wsvs}
\ee
and
\be
W_T \, \equiv \, \frac{a^{\,\prime\prime}}{a} \, = \, \frac{V}{3} \ e^{\,\frac{2}{3}\, {\cal A}} \left[ 2 \ - \ \left(\frac{d \vf}{d\tau}\right)^2 \right] \ , \label{Wtvs}
\ee
which in their turn determine the corresponding spectra via the WKB amplifications factors. Here $a$ is the conformal factor of eq.~\eqref{conf_factor} and ``primes'' denote as usual derivatives with respect to the conformal time $\eta$ of eq.~\eqref{conf_time} and one is interested in the final descent, where the second exponential in eq.~\eqref{exppot2} dominates so that $V_{\vf}\approx 2\,\gamma\, V$ and $V_{\vf\vf}\approx 4\, \gamma^{\,2}\, V$.

One can verify that the ratio $\frac{W_S}{W_T}$, which determines via eq.~\eqref{WKBlim} the corresponding effects on the power spectra, equals one for the limiting speed \eqref{limspeed} and is smaller for the larger values that, as we have argued in Section~\ref{sec:climbing}, characterize the descent in the two--term potential. Actually, for $\frac{d\vf}{d\tau}$ close to the attractor value \eqref{limspeed} one can show that
\be
\frac{W_S}{W_T} \, \approx \, 1 \ - \ 18\, \frac{(1\,-\, \gamma^{\,2})^4}{(2-3\, \gamma^{\,2})} \ \left( \frac{d \vf}{d\tau} + \frac{\gamma}{\sqrt{1\,-\, \gamma^{\,2}}} \right)^2 \ ,
\ee
and therefore the effect presents itself symmetrically about the attractor speed and is in principle sizable for the relatively small values $\gamma <\frac{1}{\sqrt{3}}$ that grant an eventual slow roll.

In the numerical investigations described in Section~\ref{sec:numerics}, scalar and tensor perturbations were nicely behaving along these lines. There we had also reconsidered our findings in terms of spectral running as the inflaton comes close to the attractor, and the arguments of this section provide complementary information to the same effect. In Appendix~\ref{a3} we shall present some analytic models for $W_S$ whose exact solutions capture the qualitative features displayed by the power spectra of figs.~ \ref{spectra} and \ref{spectra10}, aside from their oscillatory patterns.

\vskip 24pt


\scs{Conclusions}\label{sec:conclusions}


In this paper we have analyzed the imprints on the CMB of a non-standard phase of cosmological evolution that manifests itself before the onset of slow-roll inflation in a class of string models where SUSY is inevitably broken at the string scale. In this phase, the only consistent solution for the scalar field which connects to an initial curvature singularity (the cold Big Bang), forces it to be initially climbing up an exponential potential  \cite{dks}. In the simplest two--exponential potential \eqref{exppot2}, this scalar field quickly reverses its motion and begins to slow-roll inflate, even as it remains rather displaced off its attractor solution in field space. This scenario was shown in \cite{dks} to be inevitable for steep exponential potentials, whose logarithmic slope reaches (in absolute value) a critical limit that, interestingly, is attained both in the simplest ten--dimensional string model with ``brane SUSY breaking'' \cite{sugimoto} and in the four dimensional KKLT scenario \cite{kklt} with an F-term uplift \cite{uplift}. In models where a steep potential term appears in combination with another slow-roll exponential, the climbing phenomenon is followed by a subsequent slow-roll inflationary regime.  Whereas we discussed the climbing phenomenon within a specific context, these results reflect a more general Kasner--like behavior that can be exhibited working directly in cosmological time close to the initial singularity, proceeding along the lines of the BKL analysis in General Relativity \cite{cezar}. In this more general setting, one can see that the climbing phenomenon continues to occur in more general scalar potentials containing steep portions that constrain the behavior right after the singularity. Higher--curvature string corrections might also be compatible with the climbing phenomenon, at least in special classes of constructions, but we do not have conclusive arguments to this effect at the present time.

We have shown that if such a climbing phase had occurred within six to seven $e$--folds to when the largest scales accessible to current observations exited the horizon, specific imprints would manifest in the CMB power spectra. In addition to long--wavelength \emph{suppression} and \emph{enhancement} of the \emph{scalar} and \emph{tensor} primordial spectra respectively, both are generically imprinted with superposed oscillations. We provided some semi--analytic arguments for these features based on the in--in formalism for computing cosmological correlation functions, on a WKB analysis of fluctuation modes and on special analytic models for the MS potentials $W_S$ and $W_T$. Further indications were provided by the spectral running implied by inflating off the attractor. This modification of power at long wavelengths translates directly into a reduction in low multipole moments of the scalar angular power spectrum, which could become more significant than cosmic variance limits if the climbing phase occurred within three to four $e$--folds of when the largest scales accessible to us exited the horizon, and in a corresponding but milder increase in low multipoles of the tensor power spectrum. It is of course difficult to argue that the suppression of the scalar power spectrum at large wavelengths is an unambiguous predictions of the climbing phenomenon, but for instance for descending solutions, an option that appears disfavored in String Theory but would nonetheless be available if the first exponential in \eqref{exppot2} were replaced by another lying below the critical slope, or in the absence of the ``hard'' exponential brought about by String Theory, the consequences for CMB observables would be far less dramatic.
In order to obtain inflation successfully in the class of models we consider, the mass parameter ${\overline M}$ defined in Section~\ref{sec:numerics} should be of order $10^{15}-10^{16}$ GeV, which is reasonably achieved in the SUSY breaking mechanism of \cite{sugimoto,bsb} that is closely linked to the string scale. In a 4d KKLT context, this scale would reflect an energy scale of the uplift, or an appropriate combination of string scale and \emph{vev} of the axion-dilaton fields in a compactification of the ten--dimensional ``brane SUSY breaking'' setup.

It is tempting to ascribe at least part of the large--angle anomalies in the CMB \cite{low_power}, and in particular the evident lack of scalar power at the largest observable wavelengths \cite{wmap7}, to a pre-inflationary dynamics of this type. We hope to have convinced the reader that the class of models that we have analyzed, well motivated by String Theory, can provide an aesthetically compelling scenario for the Early Universe that not only gives a natural impetus for inflation to begin, but can also offer potentially observable predictions for the CMB that might allow a better fit to existing data.
\vskip 24pt


\section*{Acknowledgments}


We would like to thank Cezar Condeescu, Filippo Vernizzi and Nicola Vittorio for stimulating discussions, and Robert Brandenberger and Gonzalo Palma for comments on the manuscript. We are also grateful to the CERN Theory Department, to the CPhT -- \'Ecole Polytechnique and to Scuola Normale Superiore for the hospitality while this work was in progress. The present research was supported in part by Scuola Normale Superiore, by INFN, by the MIUR-PRIN contract 2009-KHZKRX, by the ERC Advanced Investigator Grants no. 226455 ``Supersymmetry, Quantum Gravity and Gauge Fields'' (SUPERFIELDS) and no. 226371 ``Mass Hierarchy and Particle Physics at the TeV Scale'' (MassTeV), by the contract PITN-GA-2009-237920 and by the French ANR TAPDMS ANR-09-JCJC-0146. NK is supported in part by Grant-in-Aid for Research No.DC203 from Tokyo Metropolitan University. The work of SP has also been supported by funds from the CEFIPRA/IFCPAR project 4104-2.

\newpage

\begin{appendix}

\scs{Cosmological perturbations}\label{a1}

Let us briefly review the basics of cosmological perturbation theory that are required for our investigations. Our starting point is the action for gravity coupled to a canonically normalized scalar field
\equ{bact}{S \ = \ \frac{1}{2}\ \int d^4x\sqrt{-g}\, R \ + \ \int d^4x\sqrt{-g}\Bigl[-\, \frac{1}{2}\ \partial_\mu \,\phi\, \partial^\mu\,\phi \ - \ V(\phi)\Bigr]\ ,}
and we consider perturbations around the background solution for the scalar field
\equ{pp}{\phi({\bf r},\eta) \ = \ \phi_0(\eta) \, + \, \delta\phi({\bf r},\eta)\  }
and for the line element
\equ{pl}{ds^2 \ = \ a^2(\eta)\bigg[\,-\,\big(1 \,+\, 2\psi({\bf r},\eta)\big)\, d\eta^2 \ +\  \big(1 \,- \,2\, \psi({\bf r},\eta)\big)\, dx^i\, dx^i\bigg] \ .}
We work around an arbitrary background solution $\phi_0(\eta)$, $a(\eta)$ in the so-called longitudinal gauge \cite{cosmology}, where the only independent metric perturbation is $\psi({\bf r},\eta)$, with $\eta$ the background conformal time. Substituting this ansatz into the action and making the crafty field redefinition
\equ{vdef}{{\rm v}({\bf r},\eta) \, = \,  a(\eta)\left[\delta\phi({\bf r},\eta) \, + \, \frac{a(\eta)\, \phi_0^{\,\prime}(\eta)}{a^{\,\prime}(\eta)}\ \psi({\bf r},\eta)\right]\ ,}
where primes denote derivatives with respect to conformal time, and letting
\equ{zdef}{z(\eta) \, = \, a^2(\eta) \, \frac{\phi_0^{\,\prime}(\eta)}{a^{\,\prime}(\eta)} \ ,}
the second order variation of the action becomes simply
\equ{aso}{S^{\,(2)} \ = \ \frac{1}{2}\, \int d^{\,3} {x}\ d \eta \left[({\rm v}^{\,\prime})^2 \ - \ \left(\nabla {\rm v}\right)^2 \ + \ \frac{z^{\,\prime\prime}(\eta)}{z(\eta)}\ ({\rm v})^2\right]\ .}

We note that eq.~(\ref{aso}) is formally the action for a scalar field on a {\it Minkowski} background with a time dependent mass--like term. ${\rm v}({\bf r},\eta)$ is commonly known as the Mukhanov--Sasaki (MS) variable, and is related to the comoving curvature perturbation $\mathcal R({\bf r},\eta)$ according to
\equ{ccp}{{\rm v}({\bf r},\eta) \ = \ z(\eta) \, \mathcal R({\bf r},\eta)\ .}
Eq.~(\ref{aso}) is valid for any consistent conformally flat solution of the coupled system of eq.~\eqref{bact}, but it is not the only relevant form of the action for the perturbations since one could have chosen a field redefinition different from (\ref{vdef}). Thus, working directly with $\mathcal R$ and switching to cosmological time, eq.~(\ref{aso}) can be recast in the form
\equ{asz}{S^{\,(2)} \ = \ -\, \frac{1}{2}\, \int d^4 x\, \sqrt{-g}\, \Bigl(\frac{{\dot\phi}_0}{H}\Bigr)^2\, \partial_{\,\mu}\mathcal R\ \partial^{\,\mu}\mathcal R \ ,}
which would be the standard action for a massless scalar field in the given background spacetime sourced by the dynamics of $\phi_0$, were it not for the time--dependent normalization of the kinetic term. For our purposes, another convenient choice is the rescaled MS variable
\be
Q({\bf r},t) \ = \ \frac{{\rm v}({\bf r},t)}{a(t)}\ , \label{rms}
\ee
in terms of which the action can be cast in the form
\equ{canpa2}{S^{\,(2)} \ = \ -\ \frac{1}{2}\, \int d^{\,4} {x}\, \sqrt{-g} \, \partial_\mu {Q} \, \partial^\mu {Q} \ - \ \frac{1}{2}\, \int d^{\,4} {x}\, \sqrt{-g}\, \left[ V_{\phi\phi} \, +\, \Omega(t) \right] {Q}^2\ ,}
with
\begin{eqnarray}
\label{omdef}\Omega(t) &=& - \ \frac{1}{a^3}\, \frac{d}{dt} \Bigl(\frac{a^3 {\dot \phi_0}^2}{H} \Bigr) \ = \ \frac{d}{dt} \left( \frac{2V}{H} \right) \ ,
\end{eqnarray}
where the equivalence between the last two expressions follows from the background equations of motion \eqref{eqphit} and \eqref{Hphit}. Alternatively, one can recast the action in a form involving the two (not necessarily small) parameters,
\begin{eqnarray}
\label{eps}
\epsilon_\phi &\equiv& - \ \frac{\dot H}{H^2} \ = \ 3\, \frac{ \left(\frac{d \vf}{d\tau}\right)^2}{1\,+\, \left(\frac{d \vf}{d\tau}\right)^2} \ ,\\
\label{eta}
\eta_\phi &\equiv& \frac{V_{\phi\phi}}{V} \ .
\end{eqnarray}
The end result,
\equ{canpasr}{S^{\,(2)} \ = \ -\, \frac{1}{2}\, \int d^{\,4} x\, \sqrt{-g}\, \left\{ \partial_\mu Q\ \partial^\mu Q + \Bigl[\frac{2V_{\phi}\dot\phi_0}{H} \, +\, H^2\left(6\, \epsilon_\phi - \, 2\, \epsilon_\phi^2 \, +\, 3\, \eta_\phi -\, \eta_\phi\epsilon_\phi\right) \Bigr]\, Q^{\,2} \right\} \ ,}
is valid for arbitrary backgrounds. The special case where $\epsilon_\phi \ll 1$ implies that the background is instantaneously undergoing slow--roll inflation, whereas the smaller $\eta_\phi$ is, the longer this period of slow roll can be taken to last. Typically, a long inflation (defined as more than 60 $e$--folds worth) requires $\eta_\phi \ll 1$ as well. The advantage of working with the rescaled MS variable $Q({\bf r},t)$ is that it is \emph{constant on super-horizon scales, and remains well defined throughout the evolution} of $\phi_0$, including during its early climbing phase.

Finally, the equation of motion for the Fourier modes ${\widehat Q}_{\bf k}(t)$ of the operator ${\widehat Q}({\bf r},t)$ of Section~\ref{sec:powerspectrum} takes the relatively simple form
\be
\frac{d^2 {\widehat Q}_{\bf k}}{dt^2} \ +\ 3\, H \, \frac{d {\widehat Q}_{\bf k}}{dt} \ + \ \left[ \frac{{ k}^2}{a^2} \ + \ V_{\phi\phi} \ + \ \frac{d}{dt} \left( \frac{2V}{H} \right)  \right] {\widehat Q}_{\bf k} \ = \ 0 \ , \label{qeom}
\ee
which upon use of the background equations of motion, results in the equivalent form \eqref{qeomf}. We make use of the action in terms of both $v$ and $Q$ perturbations in the main body of the paper.

\vskip 36pt


\scs{Analytic models for $W_S$ and $W_T$}\label{a3}

As we have stressed in Section~\ref{sec:qma}, there are interesting analogies between the one-dimensional Schr\"odinger equation and the MS equation in the form (\ref{mseta}), provided one identifies $k^2$ and $W_S$ with energy and potential for the quantum-mechanical problem, together with a key difference, since in the latter case one is solving an \emph{initial--value} problem. This has the key consequence that the infinite barrier resulting from the eventual approach to the LM attractor crystallizes and largely amplifies initial fluctuations. Elaborating on the analogies between the two settings can provide a better qualitative grasp of the power spectra associated to various cosmological settings. In this Appendix we investigate some relatively simple analytical forms for the potentials $W_S$ and $W_T$ that suffice to capture the gross features of the power spectra in figs.~\ref{spectra} and \ref{spectra10}.
\begin{figure}
\begin{center}
  \epsfig{file=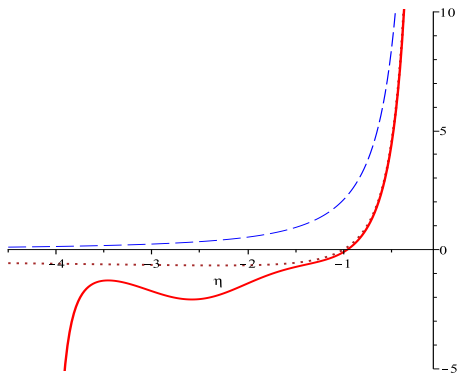, height=2in, width=2.5in}
  \caption{Comparison between attractor $W_S$ (dashed blue), two--exponential $W_S$ (continuous red) and the approximate form of eq.~(2) (dotted brown).}  \label{fig9}
\end{center}
\end{figure}

Leaving aside momentarily the Big Bang singularity, there is an interesting class of deformations of the attractor MS potential
\be
W^{(0)}(\eta) \ = \ \frac{\nu^2 \, - \, \frac{1}{4}}{\eta^2}
\ee
where
\be
\nu^2 - \frac{1}{4}\, = \, \frac{2-3\,\gamma^2}{(1-3\,\gamma^2)^2} \ ,
\ee
that captures the gross features of the actual MS scalar and tensor potentials for the two-exponential system over most of the cosmological evolution. Referring for definiteness to $W_S$, the relevant potentials read
\be
W_S = \frac{\nu^2-\frac{1}{4}}{\eta^2}\left[c \left(1+\frac{\eta}{\eta_0}\right)\ + \ (1-c) \left(1+\frac{\eta}{\eta_0}\right)^2\right] \ ,
\label{b2}
\ee
and combine the proper LM late--time behavior, a single zero and an almost flat region. These potentials depend effectively on a single interesting parameter, $c$, since $-\eta_0$ is simply the location of the zero, and have the virtue of preserving the attractor behavior of eq.~\eqref{Wsfinal} as $\eta \to 0^-$. The relevant range is $c>1$, in order that the potential tend to zero from negative values as $\eta \to -\infty$, which for one matter simulates the almost flat negative plateau present in the actual forms of $W_S$ and $W_T$. An important transition clearly takes place at $c=2$, since for $1<c<2$ the potential of eq.~\eqref{b2} lies below the LM potential at late times while the opposite is true for $c>2$. For the sake of comparison, fig.~\ref{fig9} displays the actual $W_S$ determined by the two--exponential potential \eqref{exppot2} with $\gamma=0.1$ and $\vf_0=0.5$, the corresponding attractor $W_S$ and a good approximation provided by eq.~(\ref{b2}), obtained from eq.~\eqref{b2} with $c=1.1285$ and $\eta_0=1$.

For the potentials \eqref{b2} the MS equation reads
\be
\frac{d^2\, v_k}{d\eta^2} \ + \ \left[ k^2 \ - \ \frac{(1-c)\left(\nu^2-\frac{1}{4}\right)}{\eta_0^2} \ - \ \frac{(2-c)\left(\nu^2-\frac{1}{4}\right)}{\eta_0\,\eta} \ - \ \frac{\nu^2-\frac{1}{4}}{\eta^2} \right] v_k \ = \ 0 \ , \label{b4}
\ee
or, letting
\be
\Delta \ = \ \sqrt{\left(k \eta_0\right)^2 \, + \, (c-1)\left(\nu^2-\frac{1}{4}\right)} \ , \label{b5}
\ee
\be
\frac{d^2\,v_k}{d\eta^2} \ + \ \left[ \frac{\Delta^2}{\eta_0^2} \ - \ \frac{(2-c)\left(\nu^2-\frac{1}{4}\right)}{\eta_0\,\eta} \ - \ \frac{\nu^2-\frac{1}{4}}{\eta^2} \right] v_k \ = \ 0 \ . \label{b6}
\ee

The additional substitution
\be
\rho = - \frac{\eta \Delta}{\eta_0} \ , \label{b7}
\ee
makes $\rho$ positive in the relevant range and turns eq.~(\ref{b6}) into
\be
\frac{d^2\,v_k}{d\rho^2} \ + \ \left[ 1 \ + \ \frac{(2-c)\left(\nu^2-\frac{1}{4}\right)}{\Delta\,\rho} \ - \ \frac{\nu^2-\frac{1}{4}}{\rho^2} \right] v_k \ = \ 0 \ . \label{b8}
\ee
Finally, letting also
\be
L(L+1) \, = \, \nu^2-\frac{1}{4} \ \to \ L\, = \, \nu \, - \, \frac{1}{2} \ , \quad \alpha \, = \, - \, \frac{(2-c)\left(\nu^2-\frac{1}{4}\right)}{2\, \Delta} , \ \label{b9}
\ee
eq.~(\ref{b8}) takes the standard form of the equation for Coulomb wave functions discussed for instance in Chapter 14 of \cite{A_S},
\be
\frac{d^2\,v_k}{d\rho^2} \ + \ \left[ 1 \ - \ \frac{2\, \alpha}{\rho} \ - \ \frac{L(L+1)}{\rho^2} \right] v_k \ = \ 0 \ . \label{b10}
\ee
Taking into account the unit normalization for the Wronskian in terms of $\eta$, the relevant solution for us is the combination (see eq.~(\ref{b6}), where in the flat region $\Delta^2/\eta_0^2$ plays the same role played by $k^2$ in the MS equation for the attractor)
\be
v_k \, \sim \, \frac{1}{\sqrt{\Delta}}\ \left(F_L \, +\, i\, G_L \right) \ , \label{b11}
\ee
where $F_L$ is the regular Coulomb wavefunction and $G_L$ the irregular one.
Near $\rho=0$ $v_k$ is dominated by the second term, so that the $k$ dependence of the power spectrum is captured by the limiting behavior of
\be
P_{\cal R} (k) \, \sim \, \frac{k^3}{\Delta} \, \left| \epsilon^\frac{1}{1-3\gamma^2}\, G_L(-\epsilon) \right|^2  \label{b12}
\ee
as $\epsilon$ tends to zero. This can be extracted from eq.~(14.6.7) of \cite{A_S} and from the power--like behavior of the modified Hankel function $K_{2L+1}$ near the origin, and reads
\be
P_{\cal R} (k) \ \sim \ \frac{k^3\, e^{\pi \alpha} }{\left|\Gamma(L+1+i\alpha)\right|^2\, \Delta^{2L+1}} \label{b15}
\ee
or, in terms of our original variables,
\be
P_{\cal R} (k) \ \sim \ \frac{(k\,\eta_0)^3\, \exp\left(\frac{\pi \left(\frac{c}{2}\, -\, 1\right)\left(\nu^2 \, - \, \frac{1}{4}\right)}{\sqrt{\left(k \,\eta_0\right)^2 \ + \ (c\, -\, 1)\left(\nu^2 \, -\,  \frac{1}{4}\right)}} \right) }{\left|\Gamma\left(\nu \, + \, \frac{1}{2} \, + \, \frac{i\,\left(\frac{c}{2}\, -\, 1\right)\left(\nu^2 \, - \, \frac{1}{4}\right)}{\sqrt{\left(k \,\eta_0\right)^2 \ + \ (c\, -\, 1)\left(\nu^2 \, - \, \frac{1}{4}\right)}}\right)\right|^2\, \left[\left(k \,\eta_0\right)^2 \ + \ (c\, -\, 1)\left(\nu^2 \, - \, \frac{1}{4}\right)\right]^{\nu}} \ . \label{b16}
\ee

\begin{figure}
\begin{center}
  \epsfig{file=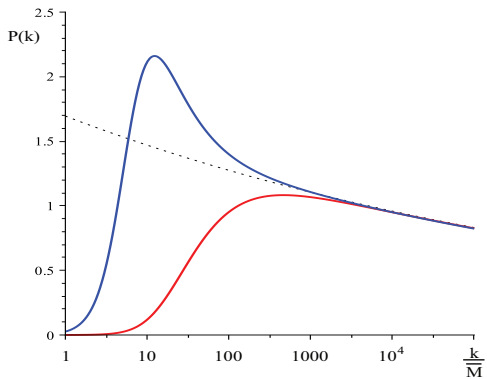, height=2in, width=2.5in}
  \caption{Analytic scalar (red) and tensor (blue) spectra vs attractor spectrum (dotted).}  \label{fig10}
\end{center}
\end{figure}

This expression combines the typical attractor behavior $P(k) \sim k^{3-2\nu}$ at large frequencies with modifications at long wavelengths that, for any given $\gamma$, depend both on $\eta_0$, which sets the scale of the phenomenon, and on $c$. The scalar power spectrum for the preferred potential $W_S$ corresponding to fig.~\ref{fig9} is the red curve displayed in fig.~\ref{fig10}, while the blue curve is the tensor power spectrum corresponding to $c=3$ and $\eta_0=3$ that results from a similar fit for $W_T$ of fig.~\ref{lmwrfig_T}. The infra-red depression present in both spectra and relative enhancement that follows it in the latter curve should be compared with the gross features of the scalar and tensor spectra of figs.~\ref{spectra} and \ref{spectra10}. One can also verify that the leading WKB approximation \eqref{WKBlim} captures the behavior of eq.~\eqref{b16} for large $\nu$.

Relatively simple analytic potentials capturing also the Big Bang singularity can be obtained at the price of turning to the less familiar and more complicated Heun functions \cite{SL}. Suffice it to say, here, that just like the Bessel and Coulomb cases describe systems with the late--time regular singular point ($\eta=0$) and one irregular singular point at infinity, the relevant confluent Heun equation possesses two regular singular points (corresponding to the initial singularity and $\eta=0$) and an irregular singular point at $\infty$.


\scs{Initial state effects}\label{a2}

In this Appendix we would like to elaborate on the effects of initial particles on the power spectrum derived in eq.~(\ref{psB}):
\equ{modf}{D_{k}\ =\ \left\{ 1 + 2 |\beta_{k}|^2 \ + \ 2 \ \cos (\delta_k + 2 \Delta_k) \ |\alpha_k \beta_k|  \right\} .}
Clearly, for isolated resonances of specific momenta, (\emph{i.e.} for an incoherent superposition of a finite number of single--particle states), the power spectrum would acquire small localized ``bumps''. As these particles are rapidly diluted by the expansion of the Universe, their contribution to the power spectrum over an observable range of wavelengths would decay exponentially\footnote{Furthermore, we take note of the observations of \cite{kundu}, wherein it was shown that not all modifications to the initial state necessarily result in a modification of the CMB two and three point correlation functions.}. In considering their presence, one should therefore justify or derive their appearance in the first place. Generically, one would arrive at Bogoliubov coefficients that would be too negligible to impart much more than a small amount of ``noise'' over the power spectrum\footnote{See \cite{tyex} for a similar observation related to an inflaton undergoing a random walk.}.

\begin{figure}[h]
\begin{center}
\includegraphics[width=6in]{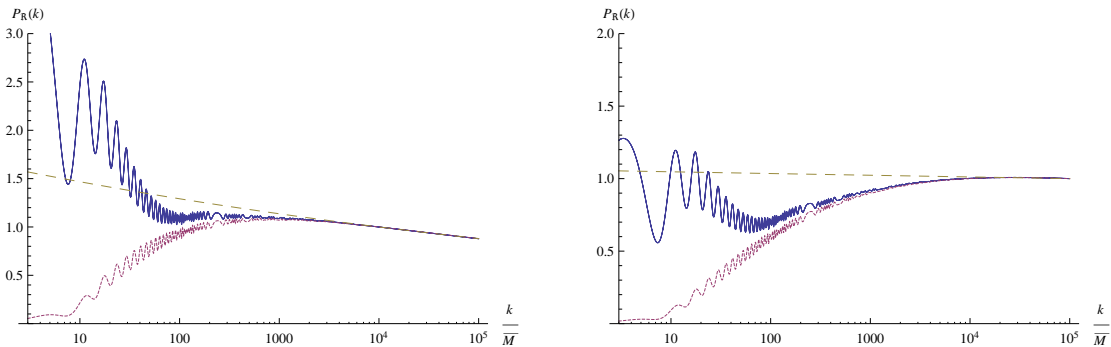}
\end{center}
\caption{Modified power spectra for $\gamma=\frac{1}{4\sqrt{6}}$ (left), and $\gamma=\frac{1}{6\sqrt{6}}$ (right). Dotted lines are the climbing scalar power spectra, dashed lines are the corresponding attractor spectra.}
\label{mod}
\end{figure}

In the scenario that we considered in the main body of the paper, the Universe begins with a ``cold Big Bang'', wherein it emerges from an initial singularity dominated by the dynamics of its lightest relevant degrees of freedom -- scalar fields representing various moduli in String Theory evolving in the potentials that have been generated for them. The fluctuations of the moduli around their (rapidly) rolling solutions can naturally be taken to be in their vacuum state, with the eventual hot Big Bang arising at the end of inflation, when the Universe reheated via the production and subsequent thermalization of Standard Model particles from the energy stored in the inflaton. However, one might ask what effects the CMB observables
would display if the fluctuations around the climbing background had been in the vacuum state natural to some other solution. Specifically, what if the Universe started out in a radiation dominated epoch, only for a climbing scalar to condense and take over the evolution? Although this is admittedly a somewhat concocted situation, it is illustrative to consider its consequences, since the initial state induced on the climbing phase would have non--vanishing occupation numbers for long--wavelength modes that one can calculate. In general, if at an initial time $t_0$ the perturbations were in the vacuum state relevant to a Universe expanding as $\left(\frac{t}{t_0}\right)^p$, computing the Bogoliubov coefficients in eq.~(\ref{psB}) via the overlaps between the wavefunctions (\ref{qlms}) with $p$ arbitrary and those with $p_c=\frac{1}{3}$, that eq.~\eqref{cosmic_climbing} associates to a climbing scalar right after the initial singularity, leads to
\be
\begin{split}
\label{btp}
\alpha_{\bf k} &= \frac{k \,\pi \,t_0}{4\sqrt{(p-1)2/3}}\ \left[H^{(1)}_\nu\left(\frac{t_0 k}{p-1}\right)H^{(2)}_1\left(-\, \frac{3t_0 k}{2}\right) + H^{(1)}_{\nu-1}\left(\frac{t_0 k}{p-1}\right)H^{(2)}_0\left(-\, \frac{3t_0 k}{2}\right) \right]\, , \\ \beta_{\bf k} &=  \frac{k \,\pi\, t_0}{4 \sqrt{(p-1)2/3}}\left[H^{(1)}_\nu\left(\frac{t_0 k}{p-1}\right)H^{(1)}_1\left(-\, \frac{3t_0 k}{2}\right) + H^{(1)}_{\nu-1}\left(\frac{t_0 k}{p-1}\right)H^{(1)}_0\left( -\, \frac{3t_0 k}{2}\right) \right] \, .
\end{split}
\ee
It is straightforward to verify that for short wavelength modes, given the standard asymptotic forms \cite{A_S}
\equ{has}{H^{(1)}_\nu(z) \sim \sqrt{\frac{2}{\pi z}} \ e^{i(z - \frac{\nu\pi}{2} - \frac{\pi}{4})} \ , \qquad H^{(2)}_\nu(z) \sim \sqrt{\frac{2}{\pi z}} \ e^{-i(z - \frac{\nu\pi}{2} - \frac{\pi}{4})}\ ,}
$|\alpha_{ k}| \to 1, |\beta_{ k}| \to 0$ as $k\to \infty$. As expected, high--frequency modes are not sensitive to the global expansion, and therefore the corresponding vacua should coincide.

Modulating the power spectrum with the factor (\ref{modf}) for the examples plotted in fig.~\ref{spectra}, initial states natural for a radiation--dominated epoch $(p = \frac{1}{2})$ rather than for the climbing phase would result in the spectra of fig.~\ref{mod}.
Therefore, the long--wavelength tail of the power spectrum would be enhanced to the point of almost nullifying the suppression induced by the early climbing phase, while significant long--wavelength oscillations are still present. As one dials up $p$ in eq.~\eqref{btp}, the effects become even larger and persist to shorter scales, although this would not correspond to any reasonable choice for the initial state of the Universe. All in all, if the Universe began in a state that is not natural for the climbing trajectory right after the initial singularity, the suppression of power at long wavelengths would be nullified, or even enhanced at the very longest wavelengths. However, such an initial state is not natural for our system, and we further remark that one would have arrived at similar long--wavelength modifications for the power spectra derived around other suitably inflating solutions (and not just the climbing solution), as is evident from the deviation from unity of eq.~(\ref{modf}) for small $k$. This is none other than the usual sensitivity of CMB observables to the pre-inflationary state if inflation lasted not much longer than it needed to\footnote{In concrete terms, one can show that the factor (\ref{modf}) can readily introduce modulations of the power spectrum of order one or larger over a span of wavelengths that would have exited the horizon up to over 4-5 $e$--folds worth of inflation. We would be sensitive to this modulation if inflation had not lasted more than 63 $e$--folds.}.

\end{appendix}

\end{document}